\documentclass[conference]{IEEEtran}

\usepackage{booktabs} 
\usepackage{flushend}
\usepackage{tablefootnote}

\usepackage{hyperref}
\hypersetup{
    colorlinks=true,
    linkcolor=red,
    urlcolor=black,
    citecolor=blue
}
\let\labelindent\relax
\usepackage{enumitem,cfr-lm}
\usepackage{fix-cm}    
\usepackage[referable,flushleft]{threeparttablex}
\renewlist{tablenotes}{enumerate}{1}
\makeatletter
\setlist[tablenotes]{label=\tnote{\alph*},ref=\alph*,itemsep=\z@,topsep=\z@skip,partopsep=\z@skip,parsep=\z@,itemindent=\z@,labelindent=\tabcolsep,labelsep=.2em,leftmargin=*,align=left,before={\footnotesize}}
\makeatother

\usepackage{amsmath,wasysym}
\usepackage{setspace}
\usepackage{amssymb}
\usepackage{listings}
\usepackage{xcolor}
\usepackage{color}

\usepackage{breakcites}
\usepackage{listings}
\usepackage[utf8x]{inputenc} 
\usepackage{framed}
\usepackage[]{caption}
\usepackage{colortbl}
\usepackage{upgreek}
\usepackage{multirow}
\usepackage{pbox}
\usepackage{array}
\usepackage{tabularx}
\usepackage{tikz}
\usepackage[percent]{overpic}
\usepackage{interval}
\usepackage{amstext} 
\usepackage{multirow}
\usepackage{bold-extra}
\usepackage{multirow}
\usepackage{bold-extra}
\usepackage{lipsum} 
\usepackage{float}
\usepackage{makecell}
\usepackage[T1]{fontenc}
\usepackage{times}
\usepackage{amsmath}
\usepackage{dcolumn}
\usepackage[T1]{fontenc}
\usepackage{times}
\usepackage{amsmath}
\usepackage{dcolumn}
\usepackage{ragged2e}
\usepackage{cleveref}
\crefname{section}{§}{§§}
\Crefname{section}{§}{§§}
\crefformat{section}{§#2#1#3}
\usepackage[T1]{fontenc}
\usepackage{bmpsize}
\usepackage{subfig}
\def\c{\textbullet} 

\usepackage{tikz}
\newcommand\encircle[1]{%
\tikz[yshift=-2pt] 
   \node (X) [draw, shape=circle, inner sep=0, scale=0.5, fill=black, text=white] {\strut #1};}

\usepackage[disable]{todonotes}

\tikzset{mynode/.style={draw = yellow,circle,inner sep=1pt,font=\normalsize,anchor=south}}
\usepackage{tabu}
\usepackage{chngpage}
\usepackage{wrapfig, mwe}

\makeatletter
\def\lst@numbersymbol{}
\lst@Key{numbersymbol}{}{\def\lst@numbersymbol{#1}}
\lst@Key{numbers}{none}{%
    \let\lst@PlaceNumber\@empty
    \lstKV@SwitchCases{#1}%
    {none&\\%
     left&\def\lst@PlaceNumber{\llap{\normalfont
                \lst@numberstyle{\thelstnumber\lst@numbersymbol}\kern\lst@numbersep}}\\%
     right&\def\lst@PlaceNumber{\rlap{\normalfont
                \kern\linewidth \kern\lst@numbersep
                \lst@numberstyle{\lst@numbersymbol\thelstnumber}}}%
    }{\PackageError{Listings}{Numbers #1 unknown}\@ehc}}
\makeatother

\definecolor{darkspringgreen}{rgb}{0.09, 0.45, 0.27}
\definecolor{dartmouthgreen}{rgb}{0.05, 0.5, 0.06}

\newcolumntype{"}{@{\hskip\tabcolsep\vrule width 1pt\hskip\tabcolsep}}

{\begin{itemize}
  \setlength{\itemsep}{1pt}
  \setlength{\parskip}{0pt}
  \setlength{\parsep}{0pt}
}%
{\end{itemize}}

\definecolor{carnelian}{rgb}{0.7, 0.11, 0.11}

\usepackage[htt]{hyphenat}
\usepackage{pifont}

\begin{document}

\title{Practical Integer Overflow Prevention}

\author{Paul Muntean, Jens Grossklags, and Claudia Eckert\\
\{paul.muntean, jens.grossklags, claudia.eckert\}@in.tum.de
\\
Technical University of Munich}


\thispagestyle{plain}
\pagestyle{plain}


\maketitle

\begin{abstract}
\todo[inline]{do this really at the end}
Integer overflows in commodity software are a main source for software bugs, which
can result in exploitable memory corruption vulnerabilities and may eventually contribute to powerful
software based exploits, \textit{i.e.,} code reuse attacks (CRAs).
In this paper, we present \textsc{IntGuard}, a symbolic execution based tool that can repair integer overflows 
with high-quality source code repairs. Specifically, given the source code of a
program, \textsc{IntGuard} first discovers the location of an integer overflow error by 
using static source code analysis and satisfiability modulo theories (SMT) solving.
\textsc{IntGuard} then generates integer multi-precision code repairs 
based on modular manipulation of SMT constraints as well as an extensible set of customizable code 
repair patterns.
We evaluated \textsc{IntGuard} with 2052 C programs ($\approx$1 Mil. LOC) available 
in the currently largest open-source test suite for C/C++ programs and with a benchmark containing large and complex programs. 
The evaluation results show that \textsc{IntGuard} can precisely (\textit{i.e.,} no false positives are accidentally repaired), 
with low computational and runtime overhead repair programs with very small binary and source code blow-up. In a controlled experiment, 
we show that \textsc{IntGuard} is more time-effective and achieves a higher repair success rate than manually generated code repairs.

\end{abstract}


%
%

\section{Introduction}
\label{intro}
\todo[inline]{do this really at the end, see what parts can be reused w.r.t. issa15 paper} 
\todo[inline]{set the stage} 
\todo[inline]{What is the problem?}  
\todo[inline]{Where are the other solutions lack?}  
\todo[inline]{How do we address the problem?} 
\todo[inline]{What is our insight?} 
\todo[inline]{Why do we implemented our checker as an Eclipse plug-in?}  
\todo[inline]{What are the benefits of using symbolic execution in this area?}  
\todo[inline]{How do you envision your tool to be used? It should be used off-line and online by the developer in order to assist them in order to pinpoint errors. Mention this}  
\todo[inline]{what is the purpose of measuring the runtime?}  
\todo[inline]{The remainder of the paper is organized as follows.}  
\todo[inline]{How does our approach improve the state-of the art of int. ov. patching approaches?}  
\todo[inline]{What is our problem description, what is the problem?}  
\todo[inline]{What is our technique, our new insight?}  
\todo[inline]{What is the goal, why do we test on a benchmark and not on real programs?}  
\todo[inline]{Be careful about false positive claims.} 
\todo[inline]{Use the motivating example in the technical description.} 
\todo[inline]{C modeling in prior works. What did other do? Why are we better than others?}  
\todo[inline]{How is our tool better than other tools?} 
\todo[inline]{Is our tool classifying benign from exploitable integer overflows? no.} 
\todo[inline]{What about the detection of previously unknown int. overflow? How do we address this issue?}  
\todo[inline]{What about systems that work on binary code, what are their advantages and disatvantages related to our work.} 
\todo[inline]{What are our contributions?} 
\todo[inline]{What about the detection of underflow/signedness and truncation, why do we focus only on the overflow. How can these other categories addressed by our too?} 
\todo[inline]{We annalyzed 2052 programs and our analysis is intra-procedural flow sensitive. State this from the beginning.} 
\todo[inline]{What about approximate path coverage? What do you mean with this in the section 4.2?} 
\todo[inline]{Avoid using not explained abreviations, CFG, CDT, etc.} 
\todo[inline]{Talk about the efforts of scalling to real programs. State why this is not our goal at first and what are the advantages and the dissatvantages of such tools. 
State what is our advantage and what others do.} 
\todo[inline]{State why the blow-up of the binary in some cases matter and why this is important in some scenarios.}  
\todo[inline]{One of the main advantage of our tool and techniques is that it can help the user to detect int. overflows during development phase thus avoiding some scalability
issues which are usually encountered with real systems which have a higher complexity and thus make some static analysis methods easily reach to their limits.} 
\todo[inline]{For example the checker could be used in future as running all the time in background as the user types. The information can be cached and used afterwards
when the program gets bigger and bigger.} 
\todo[inline]{sections have to be added here.}  
\todo[inline]{State how the patching of integer overflows influences the prevention of CRAs. Explain the needed requirements and how this influence such and cra.} 



\begin{figure}[t!]
\centering
 \makebox[\columnwidth]{
  \includegraphics[scale=.7]{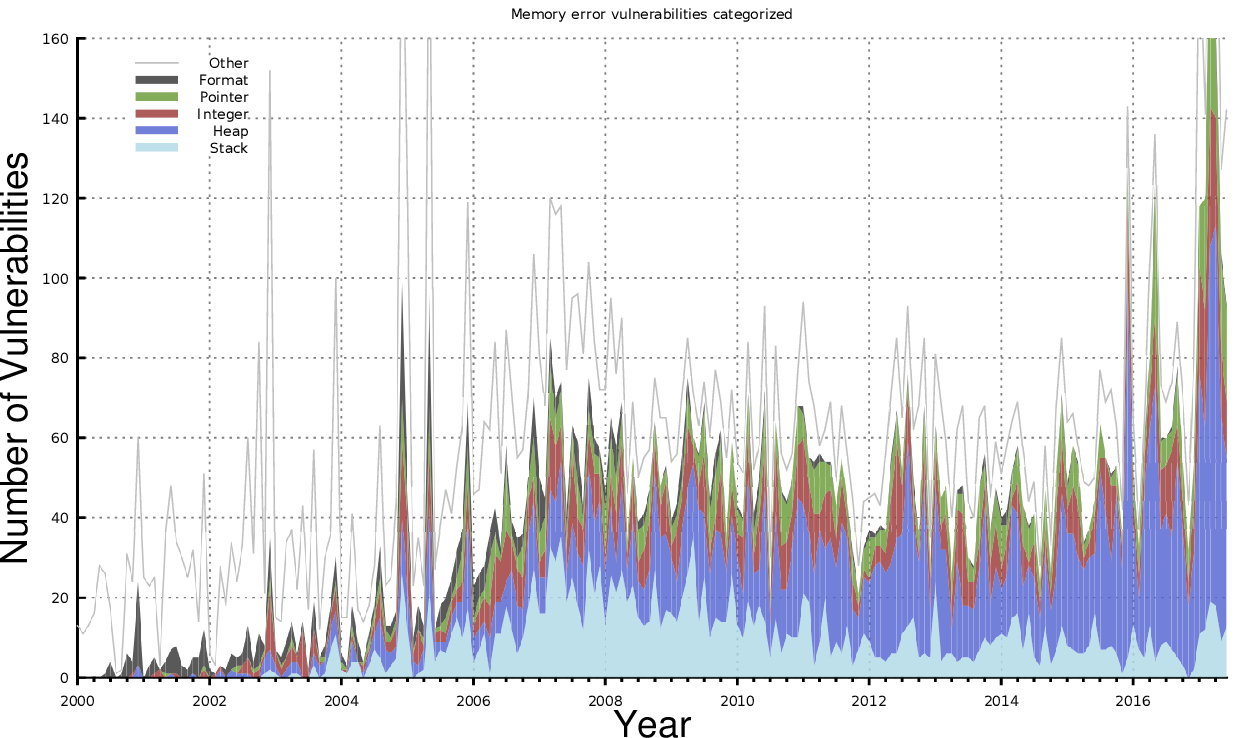}}
\vspace{-.5cm}
\caption{Integer related vulnerabilities reported by U.S. NIST.}
\label{Integer overflows reported over the last 10 years.}
\vspace{-.5cm}
\end{figure}

Integer overflows are challenging to repair and thus resulting memory corruptions are hard to avoid for at least four main reasons.
First, integer overflows are of several types~\cite{brumley:rich}:
overflow~\cite{cwe190:overflow},
underflow~\cite{cwe191:underflow},
signedness~\cite{cwe195:signed},  
truncation~\cite{cwe197:truncation}, and illegal uses of shift operations.
Second, integer overflows are typically exploited indirectly~\cite{brumley:rich}, through stack or heap overflows, which can 
lead to buffer overflows. In contrast, buffer overflows can be exploited directly or indirectly.
Typical integer related vulnerabilities (see Figure~\ref{Integer overflows reported over the last 10 years.} for more details) lead to the following exploit types:
Denial of Service (DoS),
arbitrary code execution,
bypass of an upper bound sanitization check, logic error, and array index attacks caused by a vulnerable integer value.
Third, integer overflows are hard to detect as these vulnerabilities can reside deep in the program, and manifest only for certain input types.
Finally, required integer overflow repair guarantees are hard to achieve. Reasons include that:
the location~\label{my:guarantees} where the generated integer overflow repair should be inserted is hard to determine;
generating the right repair format such that the repair is compilable, syntactically correct and does not change the program intended behavior is not trivial;
the overhead introduced by a repair is quite high if inserted na{\"i}vely in all locations prone to integer overflow (particularly in hot code such as loops);
deciding if an integer overflow was correctly removed after inserting a repair is not well addressed;  
determining whether an integer overflow manifests itself across multiple integer precisions is not trivial; and 
assessing if the integer overflow is intended or unintended is difficult to decide and less researched.

Recent solutions can be used to repair the integer overflows statically in order to avoid them during runtime. For example, Sift~\cite{sift}, TAP~\cite{tap},
SoupInt~\cite{soupint}, AIC/CIT~\cite{coker:integers}, and CIntFix~\cite{CIntFix} rely on a variety of techniques
depicted in detail in Table~\ref{tab:truthTables:bla}. In addition, several commercial tools~\cite{commercial:tools} such as Coverity Static Analysis, CodeSonar, 
Klocwork Insight, Polyspace, and Infer are available. However, since information about the detailed internals are mostly not public, we do not focus on these approaches in our work. 
Importantly, from the non-commercial tools only TAP first detects the integer overflow and then attempts to remove it by repairing it. Thus, the other tools may 
insert repairs that were not first categorized as genuine integer overflows, or not propose repairs at all and as a consequence may change the intended program behavior.
Further main challenges hindering wide adoption of these tools are their low precision w.r.t. true positive detection, 
their high runtime performance overhead (\textit{i.e.,} mostly the runtime based tools) and that the generated repairs provide only few to no repair guarantees.
Thus, these repairs have to be considered unsafe and of questionable value.

Sift~\cite{sift}---to the best of our knowledge the only currently \textit{sound} program repair tool---is a static input filter generation tool 
that inserts input filters in the program binary for which the source 
code was previously manually annotated with source code annotations.
While making a substantial contribution to the field, this tool has several limitations confirmed by the authors (of which we list eight):
 it relies on user source code annotations that are not always available for a source code program or require a considerable annotation overhead (\textit{e.g.,} annotating large code bases), 
 upper bound source code annotations for loops are needed when the analyzed expression depends on a number of values that is not finite,
 not all types of integer overflow relevant sites are supported (\textit{i.e.,} only memory allocations and block memory copy sites),
 for some types of applications (\textit{i.e.,} web servers, etc.) with no available input format specification (\textit{i.e.,} no image or video files) Sift cannot be applied,
 the tool cannot guarantee that no unwanted program behavior is introduced since the filters may remove any integer overflow - even intended ones, 
 does not support multi-precision integer overflow repairs, 
 annotated stub standard C library functions need to be provided upfront for functions that influence the computed symbolic condition (if not provided, the filter will not be generated), and finally, 
 the tool cannot help to avoid integer underflows even though it is straightforward to extend the algorithm to consider underflows as well.


In this paper, we introduce \textsc{IntGuard}, an automated tool for the generation of useful and high-quality integer overflow repairs for C source code programs.
\textsc{IntGuard} is non-intrusive and generates multi-precision sound integer overflow repairs
and is intended to be used by programmers throughout their daily routine~\cite{early:repairs}. Further, programmers should serve as the final link in the repair insertion 
decision chain in order to account for very cautious vendors~\cite{vcc}, which allow a code patch to be applied only after a thorough review.
\textsc{IntGuard} is based
on a conservative novel technique for automated generation of repairs for C source code relying on satisfiability modulo theory (SMT) constraint solving. Note that \textsc{IntGuard}s goal is not to classify
integer overflows, which can result in a memory corruption (\textit{e.g.,} if it is used to determine the size of a heap-buffer).
Instead, \textsc{IntGuard} can determine if an integer overflow is exploitable by performing a data flow analysis between a certain 
program source (\textit{i.e.,} user input) and a program sink (\textit{i.e.,} the source code location where the bug manifests).
Our novel technique relies on several building blocks.
First, we apply static symbolic intra-procedural and context-sensitive source code analysis for error detection and repair. Second, we propose a novel automated C source code repair generation algorithm based on SMT solving and code repair stubs.
More precisely, our integer overflow repair generation technique is based on modular manipulation of SMT constraints based on
deletion and insertion of new SMT constraints with the goal of generating new SMT constraint systems used for assessing if the generated 
repair satisfies some of the imposed repair requirements. And finally, we develop a semi-automatic repair insertion technique, which 
satisfies our previously mentioned repair requirements (\textit{e.g.,} preserve correct program behavior).

The \textit{soundness definition} we employ in this paper is as follows. 
The automatically generated code repairs are not program input dependent (\textit{i.e.,} SMT constraints and the Z3~\cite{z3:smt} solver are used).
The underlying integer overflow detection checker generates \textit{no} false positives (\textit{i.e.,} every detected integer overflow is a genuine 
integer overflow, each program loop is unrolled up to 15 times, this value is customizable).
A program repair removes a previously detected integer overflow on all program execution paths, which reach the integer overflow location, without 
inserting unwanted program behavior (\textit{i.e.,} only the integer overflow is removed).

Next to sound repairs, \textsc{IntGuard} fulfills the following two important objectives. 
First, \textsc{IntGuard} addresses four listed limitations of Sift: 
no source code annotations are needed, 
it can be applied to all types of C/C++ source code based programs, 
the tool provides support for multi-precision repairs (\textit{i.e.,} five types of multi-precision repair types are supported), 
and the repairs can be used to avoid integer underflows as well.
Second, the generated repairs are syntactically correct (and thus compilable, checked by code recompilation), and do not crash the program after insertion (\textit{i.e.,} do not introduce any unwanted program behavior). 
Finally, \textsc{IntGuard} checks for at least 54 types of integer overflows (see NSA's Juliet~\cite{juliet:test} test suite). 
Importantly, while repairs are generated fully automatically in an efficient fashion, the repair insertion supports a human-in-the-loop approach, but can also be fully automated, if required.
To the best of our knowledge, there are currently no other techniques for repairing integer overflows in C source code, which satisfy all above stated guarantees, and as such our technique improves 
the state-of-the-art by statically inferring sound integer overflow repairs. 
Furthermore, we provide the first static symbolic execution based technique for automatic generation of C source code repairs. 
Similar to our work,~\cite{muntean:buffer} use a symbolic execution based technique to repair buffer overflows in C source code.

In our evaluation, we analyzed 2052 C programs representing $\approx$1 Mil. LOC and have shown that~\textsc{IntGuard} repairs 
more programs on the same test suite across multiple integer precisions than CIntFix~\cite{CIntFix}, while inducing only
around 1\% computational overhead and ($\le$2\%) runtime performance overhead, respectively.
Note that CIntFix and AIC/CIT~\cite{coker:integers} induced 16\% and above 30\%, runtime overhead, respectively.
The program binary and source code blow-ups of the repaired programs are insignificant ($\le$1\%). 
Further, we construct a mini-benchmark to assess the precision and efficiency of our tool w.r.t. large and complex programs.
In the evaluation, we showed that \textsc{IntGuard} scales to large programs and that our tool is efficient w.r.t. bug detection and repair generation.
In contrast, to other similar tools which only aim to achieve scalability w.r.t. large programs we also search to detect and repair large programs in a systematic way (\textit{i.e.,} we conduct benchmark experiments to contrast and compare false positive and true positive rates on large software).
We further present several \text{in-the-wild} integer overflow types that can potentially be avoided if \textsc{IntGuard} is used throughout programmers' daily routines.
Taken together, these contributions qualify \textsc{IntGuard} for real-world applicability.

In summary, we make the following contributions: 
\begin{itemize}[leftmargin=.3cm]


\item  We designed a novel sound integer overflow C source code repair generation technique.
  

 \item We implemented \textsc{IntGuard}, a tool usable for detection\footnote{Demo movie integer overflow detection. \url{https://goo.gl/uNvdRp}} and 
 repair\footnote{Demo movie integer overflow repair. \url{https://goo.gl/912Jux}} of integer overflows.


 \item We provided an open-source repair tool that can be used to repair integer overflows across multiple integer precisions.

  \item We experimentally demonstrated that \textsc{IntGuard} is superior to other state-of-the-art integer overflow repair tools.
       
 \item We evaluated \textsc{IntGuard} thoroughly and show that it scales to large programs and its repairs are useful.
%
 
 \item We experimentally showed that \textsc{IntGuard} is more time-effective than manual repair generation and insertion.
 
 \item We responded in our evaluation to the call for more qualitative oriented measurements (see metrics and measurements section in~\cite{herley:security:science}) rather than mostly quantitative assessments.
\end{itemize}

The remainder of this paper is organized as follows.
In \cref{example} we present a motivating example, and
in \cref{background} we give the technical background need to understand the rest of this paper, while 
in \cref{design} we present the design of \textsc{IntGuard}. 
In \cref{implementation} we highlight brief implementation details, and 
in \cref{experiments} we evaluate \textsc{IntGuard}, while
in \cref{discussion} we compare \textsc{IntGuard} with other tools and discuss its limitations.
In \cref{related} we introduce the the related work, and 
in \cref{future} we give some future research directions.
Finally, in \cref{conclusion} we conclude this paper.

\section{Motivating Example}
\label{example}

\begin{wrapfigure}[16]{r}{.5\columnwidth} 

    {\vspace{-.35cm}
    \hspace{-.5cm}
\includegraphics[]{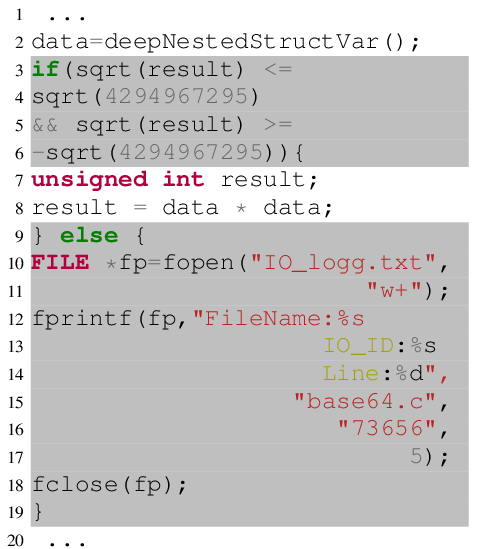}}
\vspace{-.45cm}
\captionof{figure}{Integer overflow repair.}
\label{Different types of generated integer overflow repairs.}
\end{wrapfigure}

In the following, we briefly preview our working example to demonstrate automatically generated repairs. 
Consider for this purpose the code snippet depicted in Figure~\ref{Different types of generated integer overflow repairs.}.
This code snippet was extracted from a larger program contained in~\cite{juliet:test}. Within this code example we assume that the value returned by the function \texttt{deepNestedStructVar()} and 
assigned to the variable \texttt{data} is larger than 4294967295.
Next, by multiplication of the variable \texttt{data} with itself (see code line 8), an integer overflow will be generated at program line number 8, 
since the variable in which the result of the multiplication \texttt{result} cannot hold the result of the multiplication. 
This is the case
because the variable \texttt{result} is of type \texttt{unsigned int} and this variable can not can store variables which are larger than 4294967295.
The integer overflow at line number 8 can be avoided by inserting the gray shaded code, which we assume that during the integer overflow manifestation 
was not present in the code snippet. Note, that the \texttt{sqrt()} operation accepts also variables having \texttt{double} type and when a
\texttt{long} is converted to a \texttt{double} we can have precision lost. However, for this example we consider that \texttt{sqrt()} accepts 
only integer parameters types.



\section{Technical Background}
\label{background}
This section provides the necessary background in order to understand the rest of this paper.
We start in \cref{Characteristics of Integer Overflows} by presenting the characteristics of integer overflows, and 
in \cref{Integer Overflow Related Problems.} we briefly list several integer related problems, while
in \cref{Why does it matter to avoid integer overflow based Memory Corruptions?} we present why it is important to avoid 
integer overflows. Finally, in \cref{Repair Techniques for Integer Overflow Memory Corruption} we highlight how integer overflows can be repaired, 
and in \cref{How to Prevention an Integer Overflow based Memory Corruption?} we describe how integer overflows can be detected.

\subsection{Characteristics of Integer Overflows} 
\label{Characteristics of Integer Overflows} 
Integer overflows can be classified as:
(1) intentional or unintentional, and 
(2) malicious or benign. 
An integer overflow manifests when the program gets some user-supplied
input and next the input value is used in an arithmetic operation to trigger an integer overflow.
Finally, a smaller than expected value is supplied to the memory 
allocation function and as result a smaller than expected memory will be allocated.
Deciding between the types of integer overflow related problems is rather difficult and 
a lot of research has been invested in the last years into this classification. 
A reason for this categorization effort is the desire to be able to differentiate between hard to find integer overflow errors that are more  
or less prone to exploitability w.r.t. the effort that has to be invested in order to find and repair them.

\subsection{Integer Overflow Related Problems}
\label{Integer Overflow Related Problems.} 
There are several integer overflow related problems which we will next list and briefly describe.
CWE-191, integer underflow (wrap or wrap-around)~\cite{cwe191:underflow}, 
the result of multiplying two values with each other is less than the minimum admissible integer value due to the fact that the product subtracts one value from another.
CWE-192, integer coercion error~\cite{cwe192:coercion},
manifests during bad type casting, extension or truncation of primitive data types.
CWE-193, off-by-one error~\cite{cwe193:offbyone},
during product calculation/usage an incorrect maximum/minimum value is used which is 1 more, or 1 less, than the correct value.
CWE-194, unexpected sign extension~\cite{cwe194:sign},
an operation performed on a number can cause that it will be sign extended when it is transformed into a larger data type.
CWE-195, signed to unsigned conversion error~\cite{cwe195:signed},
a signed primitive is used inside a cast to an unsigned primitive can produce an unexpected result if the value of the signed primitive can not be represented using an unsigned primitive.
CWE-196, unsigned to signed conversion error~\cite{cwe196:unsigned},
an unsigned is used inside a cast to a signed primitive, which can produce an unexpected value if the result of the unsigned primitive can not be represented using a signed primitive.
CWE-197, numeric truncation error~\cite{cwe197:truncation},
manifests when a primitive is casted to a primitive of a smaller size and data is lost in the conversion.
CWE-680, integer overflow to buffer overflow~\cite{cwe680},
manifests when an integer overflow occurs that causes less memory to be allocated than expected, which can lead to a buffer overflow.

\subsection{Avoiding Integer Overflows}
\label{Why does it matter to avoid integer overflow based Memory Corruptions?}
It is important to avoid integer overflow based memory corruptions since these are insidious, costly, and exploitable.
The {\textit{exploitability}} of integer overflow based memory corruptions is a well understood topic and for this reason {\textit{not very difficult to be performed}} by a skilled attacker,
for programs written in C/C++ since these programing languages are notoriously prone to integer overflow bugs. 
Code containing such a memory corruption has a \textit{{huge attack surface}} which can even be \textit{{exploited through the network}} by attackers in order to perform
CRAs ensuing serious consequences for all systems running that particular source code version. Open source code can be studied by attackers and new \textit{{integer overflow bugs 
can be detected with relative low effort}}
and even without tool support. \textit{{Zero-day integer overflow bugs}} in open source software have \textit{{disastrous consequences}} since these are easily exploitable and huge 
benefits gains can be achieved by the attackers. 
Finally, \textit{{integer overflow based vulnerabilities are traded online}}, as pointed out by Snowden\footnote{\url{https://en.wikipedia.org/wiki/Edward_Snowden}} and others,
and \textit{{attackers can buy integer overflow based vulnerabilities}} for a fraction of the potential damage or the achievable attacker benefit. 
As a matter of fact, we believe that, for these reasons and others not mentioned here for brevity, software should be kept as {\textit{clean}} as possible from integer overflow bugs.

\subsection{Repairing Integer Overflows}
\label{Repair Techniques for Integer Overflow Memory Corruption}
In the following, we discuss canonical repair generation strategies, and highlight relative advantages and disadvantages.

\subsubsection{Manual-Based Input Validation}
Manually written input validation checks for repairing integer overflows are challenging. First, they are error-prone and can take a significant amount of time to be inserted in large code bases. Second, they cannot guarantee that the integer overflow bug was really removed for non-trivial code locations (\textit{i.e.,} multi-dependent control flow based source code locations).
Finally, they are typically not applicable across multiple integer precisions, and cannot guarantee that the intended behavior of the program is preserved. 

Further, there are multiple documented situations where an integer overflow bug was only solved after a cascading array of repairs that missed their intended goal several times. 
We dub this type of trail of repairs as \textit{trial-and-error-repairs}. Until a repair finally succeeds, the intermediate (failed) attempts may be particularly 
dangerous if the programmer 
relaxes due to a sense of false security.
The main disadvantage of such repairs is that it gives the the wrong impression that the bug was removed,
when in reality it was not. 

\subsubsection{Compiler-Based Input Validation}
Compiler based input validation checks are cheap, and fast to insert, but can be optimized away by some compilers. For example, when employing the GCC compiler some input validation checks are useless because these can be optimized away during compilation 
since the C++ standard N4296~\cite{c++:standard} specifies that integer, arithmetic and signed overflows are considered undefined behavior, and thus implementation specific. Furthermore, the GCC developers believe that the programmers should detect the overflow before an overflow is going to happen rather than using the overflowed result to check the existence of the overflow during runtime (see detailed discussion~\cite{pldi:le, gcc:discussion}). First, in some situations this is impossible, since the search space for program inputs that trigger an integer overflow is infinite. Second, as a consequence of compiler implementation specifics, some checking conditions may be removed totally, when the program is compiled with GCC in combination with specific optimization options. 
It follows that compiler based repairs do not guarantee that the repair really removed the bug. They are also not applicable across multiple integer precisions and 
do not guarantee that no unwanted behavior is introduced. In contrast, compiler based runtime checks have in general access to more specific information than static 
tools and in some scenarios this access can provide considerable benefits w.r.t. bug prevention. 
Finally, we believe that stand-alone compilers should not be the only tool to be used for repairing integer overflows during compile time. 
Instead, specialized tools can provide more guarantees and efficiency. 

\subsubsection{Symbolic Execution-Based Input Validation}
Symbolic execution-based techniques can be used to achieve more (if not most aforementioned) repair guarantees. Furthermore,
the repairs are cheap to construct and to insert. However, these techniques are based on computationally intensive analysis strategies, which if not applied in an appropriate manner, may not scale well (or at all) with large programs. We believe that these repair tools are particularly suitable, if used early-on during development by programmers, since the level of software complexity is still low.
Finally, we believe that:
(1) manually written source code repairs should be avoided and only used in \textit{easy} to address situations,
(2) compilers should not be used for repairing integer overflows since the number of guarantees which these can offer is low, and
(3) specialized tools which provide more guarantees should be used for repair generation.

\subsection{Detecting Integer Overflows}
\label{How to Prevention an Integer Overflow based Memory Corruption?}
        \label{Program path and state coverage vs. analysis techniques.}

\begin{figure}[H]
      \centering
      \includegraphics[resolution=100000, angle=0, width=.9\columnwidth]{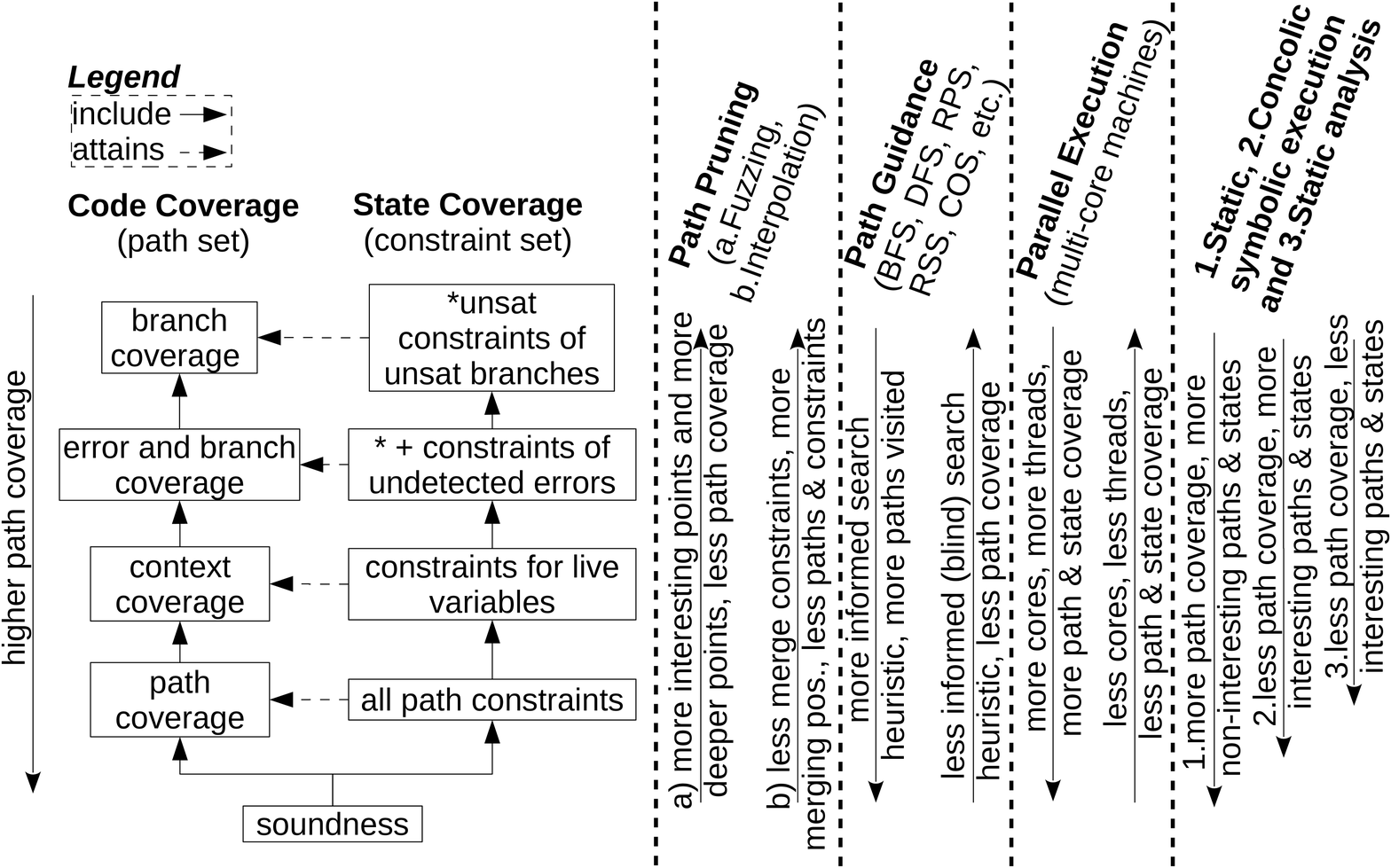}
      \caption{Path \& state coverage vs. static \& dynamic techniques.}
      \label{Program path and state coverage vs. analysis techniques.}
\end{figure}

Figure~\ref{Program path and state coverage vs. analysis techniques.} depicts the code 
coverage (\textit{i.e.,} program path coverage) and state coverage (\textit{i.e.,} symbolic variable coverage) w.r.t. the 
most used
analysis techniques to address the detection of integer overflow bugs. As far as we are aware of there is no technique which can be used for solving the problem of integer
overflow detection. Several techniques have emerged over time with more or less applicability depending on the concrete scenario in which they are applied.
We do not intend to review the advantages and disadvantages of these techniques w.r.t. to each other but rather briefly stress 
why we decided to use static symbolic execution for generation of our code repairs and briefly highlight its advantages. 
Consequently, there are several techniques which can be used to detect and repair integer overflow with more or less success. These techniques can be briefly compared 
against each other w.r.t. to program path 
coverage and state coverage. We decided to use these dimensions, depicted in Figure~\ref{Program path and state coverage vs. analysis techniques.},
since they make the most sense for our goals mentioned in the Section~\ref{intro}. In order to achieve these goals we want to 
have high path coverage and state coverage in order to have sound code repairs w.r.t. the fact that these should not change program behavior 
and integer overflow bugs should be correctly removed after applying the 
code repair. For this reason we opted for static symbolic analysis which achieves higher path coverage than concolic or purely static 
analysis techniques by visiting program paths in a depth-first search (DFS) fashion. Our static analysis technique benefits from the possibility of parallel 
execution which can be effectively used to speed up the analysis and thus more paths can 
be visited and simultaneously more states can be analyzed than in single threaded scenarios. For this purpose, we currently use a DFS strategy of path traversal which 
helps to perform a more informed search space traversal 
than without this 
technique. We plan to implement other techniques in the future. We currently perform path pruning by merging paths based on dead variables and checking satifiability of 
paths at branch nodes. This helps to drastically reduce uninteresting paths and also reduce search locations. Additionally, we check only at interesting 
locations (\textit{e.g.,} assignments) in the source code for integer overflow bugs and thus we further reduce the possible search space. 
Techniques such as fuzzing and interpolation have high priority targets on our future work agenda and we think that these can be implemented in \textsc{IntGuard} 
with ease.

\section{System Design}
\label{design}
In the following, we show in~\cref{System Overview} the system overview of \textsc{IntGuard} by reviewing its main components, and
in~\cref{Engine Capabilities} we depict the main features of our symbolic execution engine. In~\cref{Overflow and Underflow Checks}, we introduce the 
preconditions used by \textsc{IntGuard}. Next, in~\cref{Repair Generation Steps} we present the overall repair generation process based on our novel 
technique, and in~\cref{Repair Location Search} we describe how \textsc{IntGuard} can be used to find repair locations in programs, while in~\cref{Repair Process}
we discuss how repairs can be generated based on test cases. 
Finally, in~\cref{Repair Insertion Wizard2} we highlight how \textsc{IntGuard} can be used to efficiently insert code repairs into programs.


\subsection{System Overview}
\begin{figure}[H]
      \centering
      \includegraphics[scale=0.17, resolution=100000, width=.9\columnwidth]{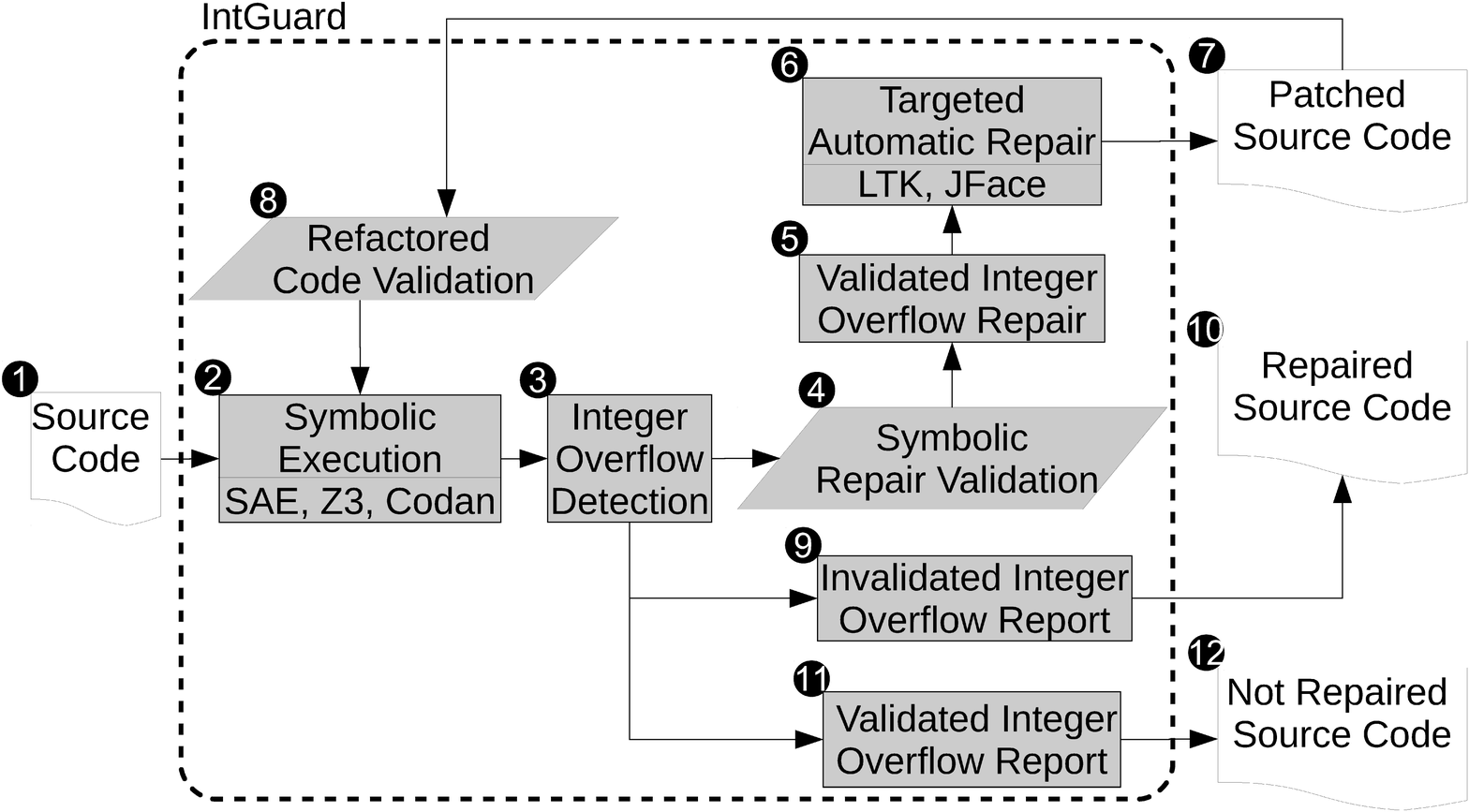}
      \caption{System overview.}
      \label{IntGuard System Overview}
\end{figure}
Figure~\ref{IntGuard System Overview} depicts the system overview of {\textsc{IntGuard}} with the corresponding components and steps
involved in obtaining the repaired programs. Initially, the {\textit{source code}} \ding{182} of the program is passed into 
the {\textit{symbolic execution}} \ding{183} engine, which we specifically developed for this purpose. 
This component can analyze C/C++ programs by first constructing a CFG for each analyzed program and second, by extracting 
program execution paths that will be symbolically analyzed. The control flow graphs and syntax trees can be shared between worker threads. 
Several workers concurrently explore different parts of a program execution tree. Each worker thread has an
interpreter 
together with a memory system model to store and retrieve symbolic variables (whose values are logic SMT-LIB equations). 

\label{System Overview}
The used symbolic execution component in Figure~\ref{IntGuard System Overview} is based on previous work~\cite{muntean:integer}.
The engine performs inter-procedural, global symbolic execution rather than a compositional analysis, and the program statement to SMT constraint translator is implemented according to the tree-based interpreter pattern~\cite{parr:patterns}. 
Unsatisfiable program branches are detected with the help of the Z3 solver~\cite{z3:smt}.
The interaction with the environment during symbolic execution is realized through {\textit{function summaries}} based 
on a C code skeleton, which specifies the name of the function, its number and types of parameters as well as return parameter type. Also, a function inside each function summary emulates the real execution of the function. This is achieved through the interplay between the interpreter and SMT statements serving as new constraints, which are added on top of the variables (similar to what happens when the real program function is executed 
during real program execution). Inside our engine, we use function summaries for network system calls.

The algorithm used for {\textit{encoding of C statements to SMT statements}} is based on a bottom-up traversal of each 
program statement located on a currently analyzed program execution path. Single static assignment (SSA) variables will be created when they are not existing for each variable encountered inside the currently analyzed program statement. Before creating a new variable, the interpreter will be queried in order to see if there is already a symbolic variable.
The C code statements are translated into SMT-LIB~\cite{barret:smtlib} equations according to the AUFNIRA logic.
The translation of Control Flow Graph (CFG) nodes into SMT-LIB syntax is performed by the translator algorithm 
which extends the CDT's abstract syntax tree (AST) visitor class according to the visitor pattern~\cite{gamma:patterns}. 
At the end, when the SMT statement creation is finished for a particular program statement, the SMT statement will be attached 
to one of the symbolic variables, which is a counterpart of one of the variables contained in the original program statement. 
During program execution traversal when an assignment statement (\textit{i.e.,} others can be also supported) was encountered, the {\textit{integer overflow detection}} \ding{184} component is notified in order to 
check if there is an integer overflow error or not. In case there was an integer overflow error, this event will be reported to 
the {\textit{symbolic repair validation}} \ding{185} component. This component is used in order to check if the negated SMT 
constraints, which were used to check for the presence of the integer overflow, could be used to invalidate the bug. In case that the result of validation is
a {\textit{validated integer overflow repair}} \ding{186}, a {\textit{targeted automatic repair}} \ding{187} can be generated. This repair has the goal of removing
the previously detected integer overflow. 

After applying the repair, we obtain the {\textit{patched source code}} \ding{188} which is represents the original code with the additional
repair that was previously added. The repaired code is again validated through the {\textit{refactored code validation}} \ding{189} which consists of
re-analyzing repaired code with the help of the symbolic execution component. Optionally, at this step the repaired code can be re-compiled in order to check if the code 
is syntactically correct.
Next, the {\textit{integer overflow detection}} \ding{184} component will be used a second time in order to validate the bug. In case the bug is invalidated, we obtain 
an {\textit{invalidated integer overflow report}} \ding{190} indicating that the bug was successfully removed and as such we get the 
{\textit{repaired source code}} \ding{191}.
In case, the integer overflow was not removed after repair insertion, then the old (or a new) integer overflow error will be 
reported, {\textit{validated integer overflow report}} {\tiny\encircle{\Large{11}}}. Then, the 
result is {\textit{not repaired source code}}
{\tiny\encircle{\Large{12}}}.

\subsection{Symbolic Execution Engine Features}
\label{Engine Capabilities}

In the following, we present the main features of our symbolic execution engine.
\subsubsection{Unrestricted Context Depth} 
Our symbolic execution engine (see \ding{183} in Figure~\ref{IntGuard System Overview}) performs an inter-procedural path-sensitive 
analysis with a call string approach~\cite{khedker:flow, sharir:pnueli}.
For each function a CFG is built. The function call context is represented by a program path leading to its call. 

\subsubsection{Loop Unrolling Trade-offs} 
Each program loop can be unrolled up to a certain depth or set to be unlimited. Currently, we unroll each loop up to 10 times.
We are aware that this incurs accuracy degradation and makes our approach to be unsound since not all 
possible program paths can be analyzed. It is also possible to leave loop unrolling unconstrained, which 
can lead in some cases to non-termination (\textit{e.g.,} endless loops) where the number of loop iterations is not known upfront.

\subsubsection{Library Calls} 
\textsc{IntGuard} can handle library calls (\textit{e.g.,} \texttt{memset}, \texttt{memcpy}, etc.) by providing upfront for each 
of this functions a stub function which models the several function behaviors:
(1) required (consumed) variables, 
(2) returned variables,
(3) operations performed on the variables of each of these functions, and 
(4) SMT constraints which model interaction of the variables used as parameters for these functions with other variables are attached to the corresponding symbolic variables. 
During static program path analysis when a library function is encountered than one of the available function library modeling functions is used by the engine interpreter.
Finally, we assume that for each analyzed program containing a library function call we have the corresponding stub function defined upfront inside our engine.

\subsubsection{Finding Program Paths} 
We use a fixed deterministic thread scheduling algorithm for running the symbolic execution.
The symbolic execution is run with approximate path coverage which uses Depth-First Search (DFS). 
During DFS, program states are backtracked and branch decisions are changed.
The loop iteration bound can be configured either to prune a path until the loop 
iteration bound is reached or to bypass the loop by avoiding the branch validation check.

\subsubsection{Automatic Slicing} 
The goal of automatic slicing~\cite{tip:slicing} is to keep the system of satisfiability checks as small as possible.
For this reason only relevant logic equations are passed to the solver for 
verification. We perform automatic slicing over the data flow in order to verify conditions on a program path 
and over the control flow to separate the analysis of different program paths.

\subsubsection{Context Sharing for Different Checkers} 
Our engine allows that multiple checkers (\textit{e.g.,} integer overflow, buffer overflow, race condition, etc.) run in parallel
during symbolic execution. The checkers are allowed to share the contexts because there is separation from the 
symbolic path interpretation. Each checker is allowed to receive notifications from a symbolic interpreter
in order to query context equations.

\subsubsection{Logic Representation} 
We use the SMT-LIB sub-logic of arrays, uninterpreted functions and non-linear integer and real
arithmetic (AUFNIRA) since this approach can be automatically decided. Pointers for example are handled as symbolic 
pointers by the engine interpreter with a target and a symbolic integer as offset formula
and outputted as logical formulas when dereferenced.

\subsubsection{Path Validation} 
The path validation is triggered at branch nodes and uses the same interface as the checkers.
For all path decisions up to the current branch the 
path validator queries the equation SMT-LIB linear system slice.
The path validator throws a path unsatisfiable exception if the solver answer is unsatisfiable;
then the symbolic execution proceeds with the next path.

\subsubsection{Satifiability Modulo Theories Solving} 
\textsc{IntGuard} uses SMT solving for three main purposes. First, for checking satisfiable and not 
satisfiable program execution paths. This helps to asses if certain program locations can be reached during 
normal program execution. Second, for checking the presence of an integer overflow additional SMT constraints are
added to SMT linear constraint system which was used to check path satisfiability in order to check for the presence of an integer overflow. Third, for 
checking if a repair removes an integer overflow additional SMT constraints are added to the SMT linear constraint system which was 
used to check path satisfiability. Finally, for solving all the SMT constraints, we use the Z3~\cite{z3:smt} 
solver because of its precision and reduced runtime overhead.

\subsubsection{ Path-sensitive Tracing of Shared Variables} 
It is possible to use shared variables between threads in a context-sensitive way. 
This is accomplished by first marking all global shared variables, and then the shared property is inferred over 
data flow constructs such as references, assignments, function call parameters and return values.

\subsubsection{ Posix Threads Support} 
We can accommodate environment functions by specifying symbolic models of library functions.
In this way, the interaction with the environment can be simulated.

\subsubsection{ Deep Nested C Structs} 
The C language program statement to SMT translator engine component is designed to recursively traverse deep nested C/C++ structures
in order to determine the real type of a field inside the struct. Accordingly, our tool does not lose precision when dealing with C structs data types.

\subsubsection{ Eclipse Extension} 
Our engine is integrated into the Eclipse CDT API as a plug-in for the following main reasons:
\textit{1)} to be able to parse C/C++ code, \textit{2)} to build a CFG
of the code \textit{3)} to allow for precise analysis of program statements with an abstract syntax tree (AST) visitor, and 
\textit{4)} to translate source code program statements to SMT constraints.

\subsection{Overflow and Underflow Checks}
\label{Overflow and Underflow Checks}
Detecting if an arithmetic operation will overflow can be 
reduced to checking that for an addition or subtraction 
of $n$-bit values the result can be stored into a $n+$1 bits precision variable. 
For a checked $n$-bit multiplication, the result should be stored in a $2n$ bits precision variable.
As~\cite{dietz:ioc} note, detecting if the result fulfills the above stated conditions is difficult.
There are several ways to detect an overflow of an operation having two signed integers $s_{1}$ and $s_{2}$.

First, we give the precondition used by the compiler based tool IOC~\cite{dietz:ioc} to avoid an integer overflow
during runtime. Our goal is to highlight the differences between our precondition checks and the one employed by IOC.
Signed addition will wrap if and only if the next expression evaluates to true.
\begin{equation*}
\begin{aligned}
  ((s_{1} > 0) \land (s_{2} > 0) \land (s_{1} > (\text{INT\_MAX} − s_{2}))) \lor \\
 ((s_{1} < 0) \land (s_{2} < 0) \land (s_{1} < (\text{INT\_MAX} − s_{2})))
\end{aligned}
\end{equation*}
IOC checks that if the above precondition evaluates to false than a failure handler will be called or 
the integer overflow prone operation will be executed.

Second, we give the preconditions used by \textsc{IntGuard}. These are extending the above precondition for other arithmetic 
operations, for integer underflows and for different integer precisions.

\textbf{Precondition 1.} The addition of two integers in which one is a variable and the other is a constant will not lead to an integer overflow or to an integer underflow if the following expression evaluates to true; $s2$ is the constant.
 \begin{equation*}
\begin{aligned}
  ((s_{1} > 0) \land (s_{2} > 0) \land (s_{1} \le (\text{INT\_MAX} − s_{2})) \land \\ s_{1} \ge (-\text{INT\_MAX} − s2))) 
\end{aligned}
\end{equation*}

\textbf{Precondition 2.} The multiplication of two integers in which one is a variable and the other is a constant will not lead to an integer overflow or to an integer underflow if the following expression evaluates to true; $s2$ is the constant.
 \begin{equation*}
\begin{aligned}
    ((s_{1} > 0) \land (s_{2} > 0) \land (s_{1} \le (\text{INT\_MAX} / s_{2})) \land \\ s_{1} \ge (-\text{INT\_MAX}/s2-1))) 
\end{aligned}
\end{equation*}

\textbf{Precondition 3.} The multiplication of two equal integers will not lead to an integer overflow 
  or to an integer underflow if the following expression evaluates to true.
 \begin{equation*}
  \begin{aligned}
   ((s_{1} > 0) \land (s_{2} > 0) \land (\text{sqrt(}s_{1}\text{)} \le (\text{sqrt(}\text{INT\_MAX}\text{)})) \land \\ 
   -\text{sqrt(}s_{1}\text{)} \ge (-\text{sqrt(}\text{INT\_MAX}\text{)}))) 
  \end{aligned}
 \end{equation*}

In contrast to IOC, if one of the preconditions evaluates to true then the operation will be performed, otherwise a failure handler will be called.
%
%
%
%
The above preconditions are used over multiple types of inputs for $s_{1}$ or $s_{2}$ (\textit{i.e.,} \texttt{RAND32()}, this is the wrapper for the C random function, \texttt{fscanf()}, etc.).
The preconditions can be applied over multiple integer precisions, meaning that \text{INT\_MAX} can take different values depending on the currently used integer precision 
in the analyzed program.
We call this value the maximum admissible upper bound value for an integer. This value is determined automatically during program analysis.
Further, the variables $s_{1}$ or $s_{2}$ can take different types: \texttt{char}, \texttt{int64\_t}, \texttt{int}, \texttt{short}, \texttt{unsigned int}. 
In contrast to IOC preconditions which will help to avoid an unconfirmed integer overflow during compile time, our preconditions will lead to code re-factorings that will surround the code location where the integer overflow was detected and confirmed.

\subsection{Repair Generation Algorithm}
\label{Repair Generation Steps}
\begin{figure}[H]
   \centering
      \includegraphics[scale=0.17, resolution=100000, width=.9\columnwidth]{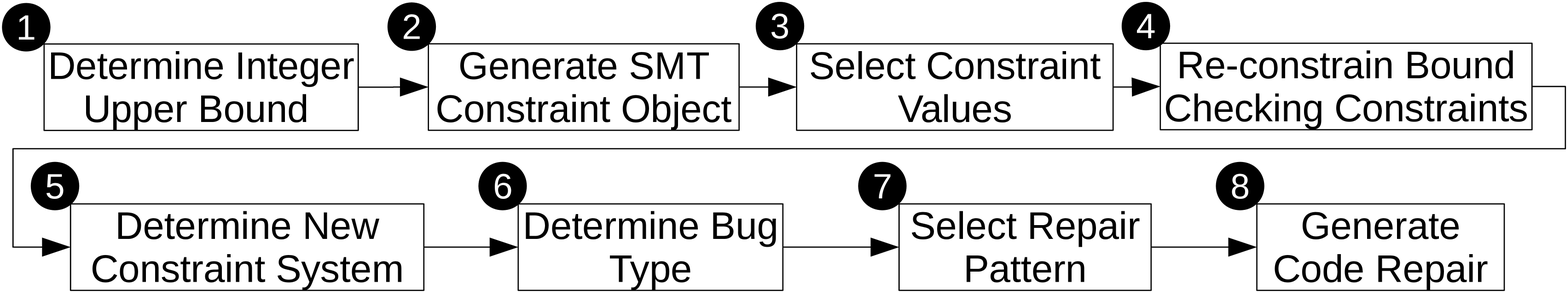}
  \caption{Repair generation algorithm main steps.}
  \label{Repair generation steps.}
\end{figure}

Figure~\ref{Repair generation steps.} depicts the eight steps of the \textsc{IntGuard} repair generation algorithm 
which are used to generate source code repairs. At the same time, these steps represent a more detailed view of the 
step \ding{191} depicted in Figure~\ref{Repair Generation Process}.

\subsubsection{Determine Integer Upper Bound}
\begin{figure}[h!]{}
    \centering
\includegraphics[]{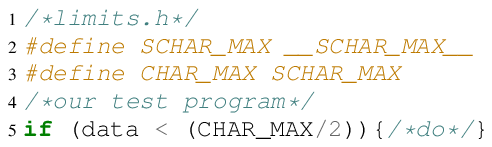}
    \captionof{figure}{Selecting upper bound value.}
    \label{Selecting upper bound value.}
\end{figure}

Figure~\ref{Selecting upper bound value.} depicts a C language code snippet extracted from an analyzed program in
which an integer upper bound value (\textit{i.e.,} \texttt{CHAR\_MAX}) is used.
In order to determine the currently used integer upper bound value in the analyzed program, \textsc{IntGuard} performs the following steps.
First, \textsc{IntGuard} parses the contents of the file located at \texttt{usr/include/limits.h} (\textit{i.e.,} Linux OS) which contains the integer upper bound values
for each integer type for the currently used OS. Note that the previous path has to be specified in advance as this may vary on different OSs.
Second, during the analyzed program path traversal \textsc{IntGuard} searches for previously in the program defined and used integer upper bound program variables.
This search is realized by comparing each declared or used variable (see if condition in Figure~\ref{Selecting upper bound value.}) name contained in the 
currently analyzed program execution path with one of the supported
integer upper bound values (\textit{i.e.,} \texttt{CHAR\_MAX}, \texttt{INT\_MAX}, \texttt{LLONG\_MAX}, \texttt{SHRT\_MAX}, and \texttt{UINT\_MAX}).
Third, in case such an upper bound value is found and used (\textit{i.e.,} inside an assignment-, addition-operation, etc.), it will be set 
to be the currently used integer upper bound value. 
In this way, the integer precision is determined for each analyzed program individually. Next, this upper bound value will be used to check for integer overflows \ding{182} and for validating generated candidate code repairs.
Finally, note that by following the above steps, \textsc{IntGuard} updates the used upper bound value automatically for each analyzed program.

\subsubsection{Clustering Bug Detection Information}
In this step, the symbolic variables and the specific constraints (\textit{i.e.,} the ones used inside the integer overflow checker) 
which were used to detect the integer overflow will be clustered and stored in 
external data structures.
\begin{figure}[h!]{}
    \centering
\includegraphics{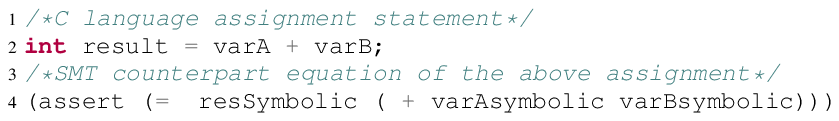}
    \captionof{figure}{C language statement and SMT counterpart.}
    \label{C language statement and SMT counterpart.}
\end{figure}

Figure~\ref{C language statement and SMT counterpart.} depicts a C language assignment statement and its SMT counterpart.
The C assignment statement depicts the addition of two variables and the result is stored in a third variable. This statement will be translated 
into the SMT constraint contained at line number 4 in Figure~\ref{C language statement and SMT counterpart.}. 
Based on the structure of the C language statement and how many symbolic variables were used to detect the integer overflow bug, 
different types of bug detection information have to be stored in external data structures for later processing. 
For the statement depicted in Figure~\ref{C language statement and SMT counterpart.} at line number 2, \textsc{IntGuard} will store:
the statement where the bug was detected,
the SMT constraint system used to detect the bug in first place,
the bug ID (to identify which bug type was detected),
the symbolic variable which was used to detect the integer overflow, and optionally other symbolic variables on which the integer overflow triggering variable depends directly.
An external data structure represents an aggregation of the symbolic variables used to detect the bug together with the particular integer overflow SMT checking constraints. 
Finally, the key reason for generating these additional data structures is to help to group relevant data together and to facilitate a better handling of this information in later steps.

\subsubsection{Select Constraint Values}
\textsc{IntGuard} selects \ding{184} the relevant SMT constraint variables based on the type of C language statement where upfront the integer overflow was detected.
\begin{figure}[h!]{}
    \centering
    \includegraphics{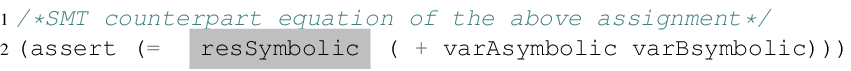}
    \captionof{figure}{Selecting relevant symbolic variable(s).}
    \label{Selecting relevant symbolic variable(s).}
\end{figure}

Figure~\ref{Selecting relevant symbolic variable(s).} depicts with gray shaded color the symbolic variable \texttt{resSymbolic}. 
This variable was selected by \textsc{IntGuard} to be further constrained in order to check if the code repair that will be generated
could remove the previously detected integer overflow bug.
Note that depending on the complexity of the analyzed statement (where the integer overflow was detected) more or less variables can
be taken into consideration in order to determine if the previously detected integer overflow would further manifest depending on how the 
selected symbolic variables were constrained. Finally, the selected variables will be used in the next step when re-constraining the bounds during the checking of SMT constraints of the SMT system (which was used upfront to detect the integer overflow).


\subsubsection{Re-constrain the Bound Checking Constraints}
After collecting all the path constraints \ding{185} for a single program execution path, \textsc{IntGuard} adds the SMT constraints which check for the presence of an
integer overflow. The presence of an integer overflow bug is indicated if for the selected SMT system the Z3 solver reports \texttt{SAT} (satisfiable, integer overflow bug present).
\begin{figure}[H]{}
    \centering
\includegraphics{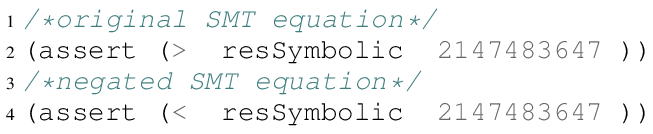}
    \captionof{figure}{The original and the newly added SMT constraint.}
    \label{Original and negated SMT equation.}
\end{figure}
Figure~\ref{Original and negated SMT equation.} depicts the original SMT equation (line 2) and a second SMT constraining equation (line 4) on the previously selected variable.
\textsc{IntGuard} re-constraints (\textit{i.e.,} through integer upper bound negation) the selected variable(s) selected from the previous step in order to determine a 
potential interval which will not lead to a second integer overflow of the symbolic variable (which previously was used to detect the integer overflow).
The goal is to determine if there will be a second integer overflow if \textsc{IntGuard} re-constraints the selected variables. For this purpose
\textsc{IntGuard} checks again in the next step if for the new SMT constraint system (see next step) it gets an \texttt{UNSAT} (unsatisfiable, no integer overflow bug present) solver reply. 
The new constrained SMT will be composed of the old SMT constraint system which was used to detect the integer overflow complemented with the re-constrained SMT equations.
If \textsc{IntGuard} gets an \texttt{UNSAT} solver reply then it determines that there will be no integer overflow if it re-constraints the selected variable(s) 
with the new constraints (\textit{e.g.,} variable range negation, etc.) and as such the integer overflow can be avoided. 
It follows, that the information collected at this step can be used to construct in a later step the final code repair.
Note that this approach can be extended to other scenarios; \textit{e.g.}, more complex constraints can be added and checked if required for other types of bugs.

\subsubsection{Determine New Constraint System}
In this step \ding{186}, \textsc{IntGuard} assembles the new SMT constraint system which will be used to determine if the previously detected integer
overflow is still present. During this step, \textsc{IntGuard} takes the constraints determined at the previous step and inserts them in the 
SMT constraint system which was used to detect the integer overflow. 
\begin{figure}[H]
\centering
\includegraphics{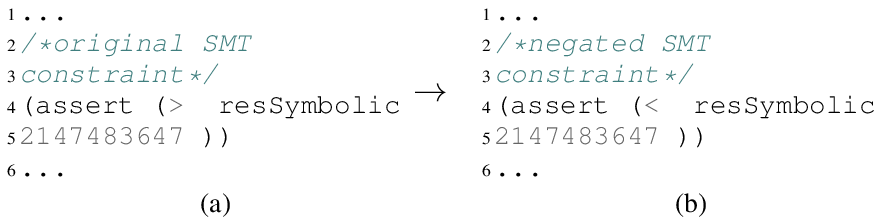}
\caption{Composing the new SMT system.}
\label{Composing the new SMT system.}
\end{figure}
Figure~\ref{Composing the new SMT system.} (a, b) depict the replacement process of the old SMT sub-system (component) with a new SMT sub-system that was determined at the previous step.
Before inserting the new constraints in the SMT system, \textsc{IntGuard} needs to 
remove the original SMT constraints that were used to detect the presence of the integer overflow. The decision 
which SMT constraints have to be removed from the
SMT system (used to detect the integer overflow presence) 
is made based on the modular aggregation of these SMT statements inside the integer overflow checker in which these constraints are put together.
More precisely, since the SMT constraints used to check for an integer overflow are added in the integer overflow checker in an incremental building 
block fashion, \textsc{IntGuard} can precisely determine which constraints can be safely removed and replaced with new ones determined at the previous step.
Next, this SMT system will be fed into the Z3 solver. In case the solver replies \texttt{SAT}, then the constraints added represent valid constraints which can be 
used to remove the previously detected integer overflow. 
Finally, this SMT constraint system will serve as confirmation that the integer overflow can be removed if certain symbolic variables are constrained in an appropriate way.
Note that these symbolic variables have counterparts, which will be inserted when assembling the final code repair.

\subsubsection{Determine Bug Type}
The integer overflow checker runs in parallel in our engine with other checkers (\textit{e.g.,} race condition checker, buffer overflow checker, etc.), which are currently 
available in our engine. Each bug checker generates a report containing an unique bug identifier 
(\textit{i.e.,} time stamp based + checker unique identifier) for each detected bug. Based on the generated bug 
identifier, \textsc{IntGuard} can determine which bug type \ding{187} it currently deals with. 
This information is extracted from the external data structures constructed at step 2.
With this information, \textsc{IntGuard} checks in the list of currently supported checkers to which checker this stored identifies belongs to.
Based on this information, \textsc{IntGuard} can determine which repair pattern can be used to repair the previously detected integer overflow. 

\subsubsection{Select Repair Pattern} 
Based on the previous determined bug identifier (see \ding{187}), \textsc{IntGuard} selects from the 
repair pattern \ding{188} pool 
the repairs suited for integer overflow repair.
\textsc{IntGuard} repair patterns consist of empty C code skeletons (stubs) where C statement parts will be replaced with
concrete values after their values have been computed as presented in \ding{185} or with other types of mathematical operations (\textit{e.g.,} division by a value).
Also, note that in some situations place holder variables
will be replaced with corresponding mathematical functions such as the square 
root function \texttt{sqrt} or other functions. In these situations, \textsc{IntGuard} does not compute the value of the function upfront but rather leaves this to be computed later
during symbolic execution analysis or program runtime. 
This offers the advantage that \textsc{IntGuard} does not need to be able to compute any possible mathematical function before
program runtime.

\begin{wrapfigure}[22]{r}{.5\columnwidth} 
    \vspace{-.35cm}
    \hspace{-.45cm}
\includegraphics{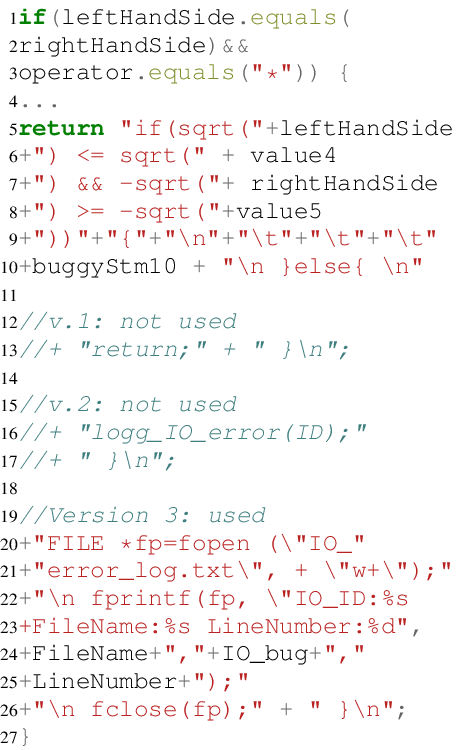}
    \captionof{figure}{Code repair pattern.}{\unskip}
    \label{Different types of repair patterns.}
\end{wrapfigure}

Figure~\ref{Different types of repair patterns.} depicts a code repair pattern used by \textsc{IntGuard} during 
integer overflow error repairing (see generated repair in Figure~\ref{Different types of generated integer overflow repairs.}).
If not noted otherwise, each code repair pattern contains C code compatible snippets (\textit{i.e.,} code in red font) interleaved with variables that 
will be inserted in the repair after the integer overflow was detected and before the repair will be inserted into the buggy program.
The code repair contains several stub variables which will be replaced with concrete variables names and values depending on the type of 
code statement containing the bug.
The code repair patterns used by \textsc{IntGuard} contain precondition checks which currently cover the preconditions 
listed in section~\cref{Overflow and Underflow Checks}. These preconditions cover multiplication of numbers and addition of variables.
At the same time the repair patterns are highly configurable and versatile. The programmer can easily change if needed for example the error handling function 
inside the repair pattern or can extend the precondition such that it captures more complex situations where for example multiple numbers are added, multiplied, divided, etc.
This can be achieved by modifying a few lines of code inside and existing pattern or by creating a new one and defining the conditions 
(\textit{i.e.,} this depends on the structure of the AST of the statement where the bug was detected) when such a pattern should be used.

\textsc{IntGuard} follows the next steps in order to select the repair pattern, which should be used for repairing a previously detected integer overflow.
First, the code statement where the integer overflow error was detected 
is divided into its components based on his AST. For example the AST components 
of a simple C statement like \texttt{int result = varA + varB;}
are: \texttt{leftHandSide=varA}, \texttt{operator=+} and \texttt{rightHandSide=varB}.
Second, a series of rules are checked on the AST of the previous C statement such as: 
(1) is \texttt{leftHandSide} equal/different than \texttt{rightHandSide}, 
(2) what type of operator do we have in the statement (\textit{i.e.,}\texttt{*}),
(3) how many components does the statement have after the \texttt{=} sign, and so on.
Third, based on these rules the repair pattern which satisfies the highest number of constraints will be selected.
Note, that each repair pattern has a list of properties (\textit{i.e.,} use when \texttt{rightHandSide=leftHandSide} and \texttt{operator=+}, etc.) attached to it that are checked against the above stated rules. This list of properties is statically 
defined when the repair patterns were manually added to the pool of available repairs.
Finally, in case there are more repair patterns that fulfill the same number of rules (\textit{i.e.,} rules match properties), \textsc{IntGuard} selects the first repair pattern 
occurring in the list. Note that if needed this approach can be updated such that all legitimate repair patterns will be proposed and for each a repair can be generated, and selected with a human-in-the-loop approach.

For example, the repair pattern depicted in Figure~\ref{Different types of repair patterns.} will be used by \textsc{IntGuard} when the \texttt{leftHandSide} equals (\textit{i.e.,} string wise comparison) with the \texttt{rightHandSide} and the
\texttt{operator} equals (\textit{i.e.,} string wise comparison) to the product operator $*$ 
(see code lines 1-3 in Figure~\ref{Different types of repair patterns.}). 

After the above checks have been performed the repair will be assembled by following the next steps.
First, \texttt{value4}, \texttt{value5} and \texttt{buggyStm10}, depicted in Figure~\ref{Different types of repair patterns.} between code lines 5-10,
are replaced with the following.
(1) the squared root of the currently selected integer upper bound value \texttt{value4 $\leftarrow$ sqrt(2147483647)}, 
(2) the negated integer upper bound value \texttt{value5 $\leftarrow$ -sqrt(2147483647)}, and
(3) the program code statement that contains the previously detected integer overflow error \texttt{buggyStm10 $\leftarrow$ int result = data * data;}). 
Second, the variables \texttt{FileName}, \texttt{IO\_ID} and \texttt{LineNumber}
depicted in Figure~\ref{Different types of repair patterns.} between code lines 23-24 are 
replaced with concrete values obtained during bug detection.

Finally, note that: (1) other code repair patterns can be selected and used based on the format of the AST of the program statement where the integer overflow was detected upfront.
(2) our approach can be easily generalized to more complex C code statements than the ones mentioned herein, and 
(3) each generated repair can be easily customized to fit to different types of integer overflow mitigation (\textit{i.e.,} error logging, calling a handling function,
see v.1 and v.2 in Figure~\ref{Different types of repair patterns.}).

\subsubsection{Generate Code Repair}
The final step \ding{189} consists in putting together the final code repair and saving it in a list of repair candidates for the previously detected integer overflow.
After all components have been inserted into the previously selected code repair pattern in \ding{188}, \textsc{IntGuard} generates a C code repair which is 
syntactically correct, can be compiled and could be further on edited after insertion (if desired). 
Next the assembled repair will be sent to the Eclipse LTK API based component of our engine, which will assemble 
the final code repair. The steps performed in this component consist in converting the code repair code into another representation based on LTK node objects, which map to the translation unit for the file in which we want to insert the repair. The LTK component
decides how to position the repair in the buggy program such that the integer overflow will not occur after the repair was inserted. The repair comes close to 
a guard around the error prone code which forbids that an integer overflow manifests during program runtime.
Finally, the repair will be passed to \textsc{IntGuard}s repair insertion component, which will create two differential views 
(\textit{i.e.,} with the repair inserted in the file containing the bug and without).

\subsection{Repair Location Search}
\label{Repair Location Search}
In order to generate the code repair, \textsc{IntGuard} needs to detect the precise location where the integer overflow
resides in the program. Next, we will present the main steps of our repair location search algorithm.
First, each program execution path is extracted from a previously computed CFG.
Second, the extracted path is traversed and path satisfiability checks are performed at branch nodes.
Third, when encountering an integer error prone code location (\textit{i.e.,} assignment statement) on the analyzed 
path, an integer overflow check is performed by notifying the interpreter.
Fourth, the notification is delegated to the appropriate checker (\textit{i.e.,} integer overflow checker) by the interpreter.
Fifth, the slice of SMT equations of the symbolic variable which overflowed is queried by the integer overflow checker and 
corresponding integer overflow satisfiability checks are added.
Sixth, the check verifies if the symbolic variable, which caused the integer overflow, can be greater (\textit{i.e.,} if true then there is an integer overflow) than 
the currently used integer upper bound value (\textit{i.e.,} \texttt{INT\_MAX}). These upper bound values are extracted from the C standard library contained in the \texttt{limits.h} file.
The lower bound is obtained by negating the currently used upper bound value.
Finally, if the solver replies \texttt{SAT} (satisfiable, integer overflow bug present) to the previously submitted SMT query, then a problem report 
(\textit{i.e.,} problem ID (unique system string), file name where the bug 
was detected and line number where the bug is located) will be created and delivered.
Finally, note that in principle all other integer overflow related problems (\textit{i.e.,} truncation, signedness)
can be detected (and repaired) by using our bug location search algorithm and repair generation technique.
Note, that only the type of error prone statement (\textit{i.e.,} most likely code statements) and the additional checking constraints
which are added by each particular checker differ w.r.t. integer- and underflow-overflow detection and repair.

\subsection{Repair Insertion Support}
\label{Repair Insertion Wizard2}
\begin{figure}[h]
\centering

\includegraphics[resolution=100000, width=.99\columnwidth]{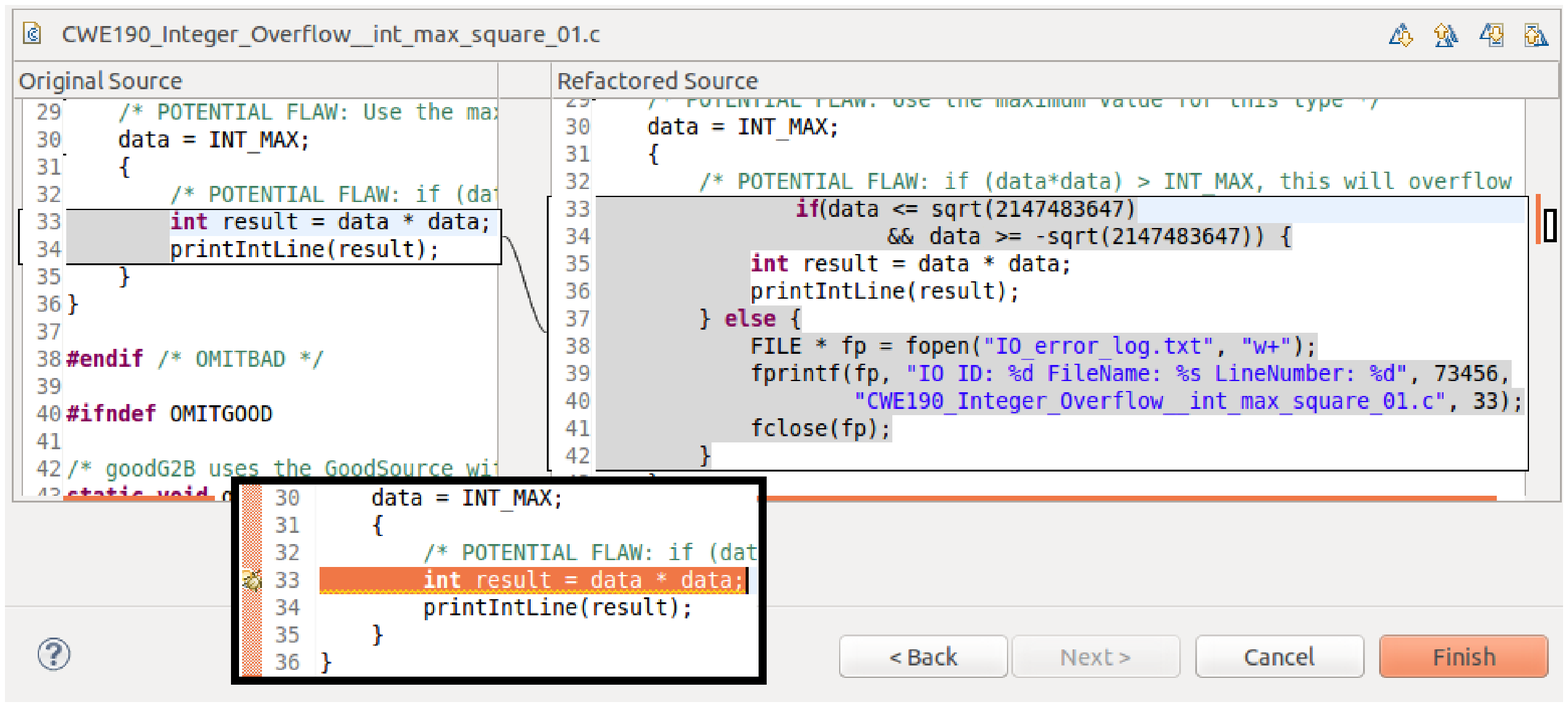}
\caption{Bug detection and repair differential view.}
\label{Repair Insertion Steps}
\end{figure}
Figure~\ref{Repair Insertion Steps} depicts how \textsc{IntGuards} GUI support for repair insertion looks like.
First, the integer overflow checker which is a sub-component of \textsc{IntGuard} places a bug marker depicted in Figure~\ref{Repair Insertion Steps}(black bordered box) with
a yellow bug icon, on the left of the C statement if at that particular code line an
integer overflow error was detected, see demo~\cite{demo1}.
Second, by right clicking on this bug marker the user can start the code
re-factoring wizard, see demo~\cite{demo2}.

The code re-factoring wizard is composed of two windows.
The first window is used to make repair type decisions (currently only in-place repairs are available). 
The second window depicted in Figure~\ref{Repair Insertion Steps}(in background)
contains a differential files view visualizing the differences between the original
file containing the bug and the modified file with the selected repair inserted.
Finally, it is possible to navigate between these two windows and if 
wanted the repair generated in~\cref{Repair Generation Steps} can be inserted by left click on the \texttt{Finish} button.

\subsection{Testcase Based Repair Generation Support}
\label{Repair Process}

\begin{figure}[H]
   \centering
      \includegraphics[scale=0.17, resolution=100000, width=.9\columnwidth]{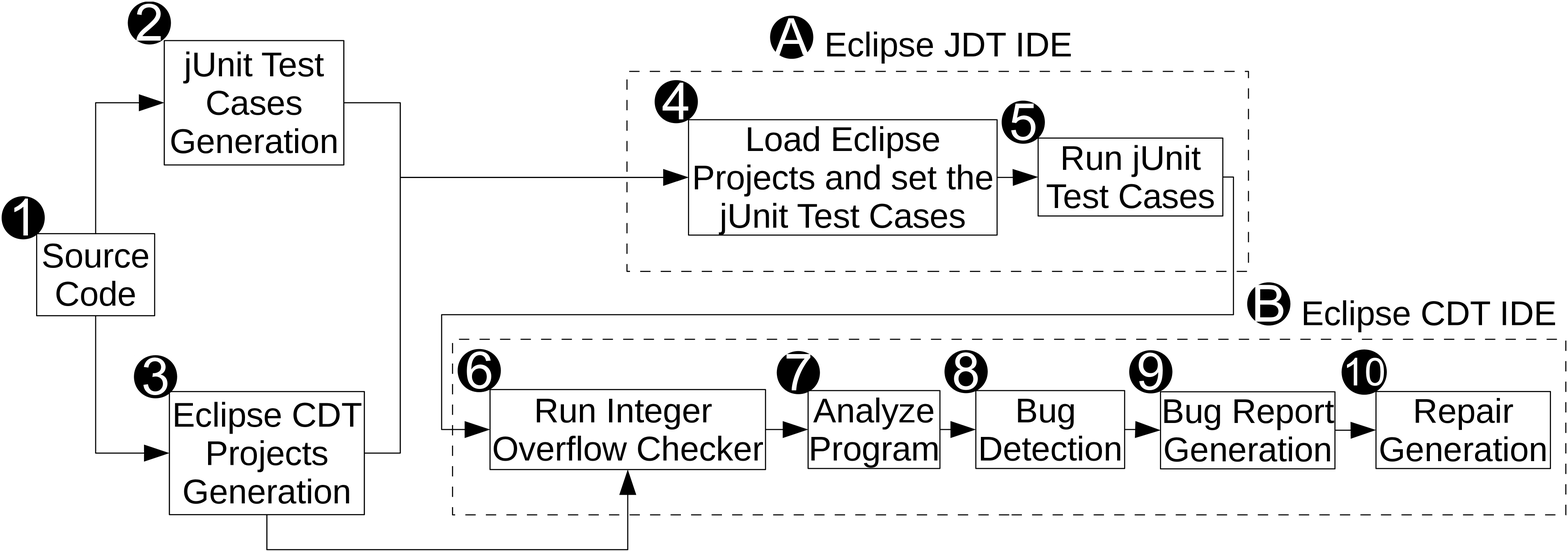}
  \caption{Repair generation process.}
  \label{Repair Generation Process}
\end{figure}
Figure~\ref{Repair Generation Process} depicts how \textsc{IntGuard} can be used to repair
buggy programs with test case based support (\textit{i.e.,} jUnit).
Note, that \textsc{IntGuard} can be used out of the box without test cases as well.
\ding{182}, represents the analyzed C source code program,
\ding{183}, highlights the automated Eclipse jUnit test case generation,
\ding{184}, depicts the automated Eclipse C/C++ projects generation,
\ding{185}, indicates the loading of the Eclipse projects and jUnit test cases, 
\ding{186}, represents the running of the jUnit test cases,
\ding{187}, depicts the running of the integer overflow checker,
\ding{188}, presents the static symbolic program analysis,
\ding{189}, depicts the bug detection,
\ding{190}, indicates the bug report generation, and
\ding{191}, highlights the repair generation steps (see section~\ref{Repair Generation Steps} for more details) for the previously generated report.
The steps \ding{185} and \ding{186} run 
inside {\tiny\encircle{\Large{A}}} (\textit{i.e.,} the Eclipse JDT IDE) and that
\ding{187}, \ding{188}, \ding{189}, \ding{190} and \ding{191} run 
inside {\tiny\encircle{\Large{B}}} (\textit{i.e.,} the Eclipse CDT IDE).
We depicted {\tiny\encircle{\Large{A}}} and {\tiny\encircle{\Large{B}}} sequentially in Figure~\ref{Repair Generation Process} since the Eclipse 
JDT IDE triggers the start of the Eclipse CDT IDE. Note that if desired the repair generation plug-in can be deployed
into the Eclipse CDT IDE such that the Eclipse JDT IDE becomes superfluous.
As a consequence, \ding{183}, \ding{185} and \ding{186} become optional
and will not be used.

\section{Implementation}
\label{implementation}
\textbf{\textsc{IntGuard}.} We implemented \textsc{IntGuard} 
based on the static analysis engine, which
we developed as an Eclipse CDT IDE plug-in. 
We followed this approach since 
(1) the Eclipse CDT API can be easily reused, 
(2) a GUI is easily obtainable, and 
(3) the obtained tool can be used in online (\textit{i.e.,} during code typing) and offline (\textit{i.e.,} after finishing code typing) mode. 
For this purpose, we used Codan~\cite{codan:laskavaia} in order to: 
(1) construct the program CFG,
(2) analyze AST of program statements, and
(3) perform bottom-up traversals by using a C program statement visitor in order to construct SMT constraints.

\textbf{Source Code Refactoring Tool.} We implemented also a graphical user interface (GUI) source code 
refactoring tool 
used 
to aid the programmer during repair 
insertion based on the Eclipse language tool kit (LTK), 
JFace and Eclipse CDT. The tool provides useful features to the programmer in order to take a more informed decision
when inserting the previously generated repair into the buggy program.

\textbf{jUnit Test Case Generator Tool.} We implemented a tool 
to generate jUnit test cases for our tested programs in order to effectively
assess the effectiveness w.r.t. false positives and false negatives of our tool. The tool generates the jUnit 
test cases fully automatically after previously providing the path to the test programs folder. The tool can automatically
infer the line number where the bug is located, the program function containing the bug, and the class name.
This information will be used to parametrize a jUnit testing function which will be used to asses if \textsc{IntGuard}
detected a true positive or false positive.

\textbf{Eclipse CDT Projects Generator Tool.} We implemented a tool 
in order to effectively assess our tool on any provided C source code program. 
The tool requires for project generation:
(1) as input the path to the main folder where the files of the program which we want to test are located, and 
(2) for each generated project, two Eclipse CDT project files having the file extensions
\texttt{.cproject} and \texttt{.project} which will be added to the project and updated with the names of the test programs which we want to 
convert to Eclipse CDT projects. The program generates new project by copying all the files of the program which we want to analyze
into an folder and the aforementioned project files. Inside these files the name of the project will be updated accordingly to the name of the program which
we want to analyze. 



\section{Evaluation}
\label{experiments}
To evaluate the usefulness of our program repairs
and \textsc{IntGuard}s scalability in general, we answer the
following research questions (RQs):

\begin{itemize}[leftmargin=.32cm]
 
 \item \textbf{RQ1:} How \textbf{effective} is \textsc{IntGuard} w.r.t. integer overflow detection (\cref{RQ4: Effectiveness w.r.t bug removal.})?
		
  \item  \textbf{RQ2:} What \textbf{false/positives negatives} ratio does \textsc{IntGuard} have (\cref{RQ2: False Positives and False Negatives})?
		
  \item  \textbf{RQ3:} What is the \textbf{repair generation overhead} of \textsc{IntGuard} (\cref{RQ2: Computational overhead.})?
		
  \item  \textbf{RQ4:} Is \textsc{IntGuard} preserving intended \textbf{program behavior} (\cref{RQ5: Preserving Program Behavior.})?
  
 \item  \textbf{RQ5:} Does \textsc{IntGuard} \textbf{scale to large programs} (\cref{Microbenchmark Correctness Evaluation})?

 \item \textbf{RQ6:} Does \textsc{IntGuard} influence the program \textbf{runtime overhead} (\cref{RQ3: Run-time overhead.})?
		
  \item  \textbf{RQ7:} Does \textsc{IntGuard} influence repaired \textbf{program size} (\cref{RQ6: Source code and binary blow-up.})?
		
  \item  \textbf{RQ8:} How easy is the \textbf{deployment} of \textsc{IntGuard} compared to other tools (\cref{RQ1: Deployment.})?
		
  \item  \textbf{RQ9:} Which CRA types can \textsc{IntGuard} help to \textbf{avoid} (\cref{RQ7: CRAs prevention})?
  
 \item \textbf{RQ10:} How does \textsc{IntGuard} \textbf{help} during bug repair (\cref{Controlled Experiment})?
		
 \item   \textbf{RQ11:} Is \textsc{IntGuard} \textbf{superior} compared to other tools (\cref{Comparison with Other Tools})?

\end{itemize}

\textbf{Preliminaries.}
Programs contained in the Juliet test suite have:
on average 476 LOC with a maximum of 638 LOC\footnote{The largest is \textit{CWE190\_Integer\_Overflow\_int\_listen\_socket\_multiply\_15}. We added the utility files contained in~\cite{juliet:test}
for each of the generated projects such that the code becomes compilable.},
real integer overflows,
exactly one true positive and several false negative integer overflows (see characteristics of~\cite{juliet:test}), and
many types of integer overflow related types.
The analyzed programs are available in the CWE-190, which is part of the currently largest open source test suite for C/C++ code~\cite{juliet:test}.
Additionally, we built a mini-benchmark containing programs having from 6K LOC up to around 20K LOC in order to show that \textsc{IntGuard} scales to large and complex 
programs and that it perfectly aligns with other static symbolic execution tools (\textit{i.e.,} KLEE).
Finally, we performed a controlled experiment with 30 participants in which we assessed the efficiency of \textsc{IntGuard}
during bug repairing.

\textbf{Experimental Setup.} We conducted the experiments on a Dell desktop with an Intel CPU Q9550 \url{@} 2.83GHz, 64-bit, 12GB RAM,
by running the Eclipse IDE v. Kepler SR1 in OpenSUSE 13.1 OS.
Similar to other symbolic execution tools, all integer overflows reported by \textsc{IntGuard} were manually reviewed to decide if they 
were false positives/negatives, the integer overflow bugs were really removed and if the inserted repair would expose the program to other vulnerabilities.

\begin{table}[h!]
 \begin{center}
 \resizebox{\columnwidth}{!}{%
\begin{tabular}{ |r|r|r|r| } 
 \hline
 \shortstack{IntGuard \\ Satisfiable Paths}          & \shortstack{IntGuard \\ Unsatisfiable Paths} & \shortstack{IntGuard Program \\ Branch Nodes} & \shortstack{IntGuard \\ Program Branches} \\ \hline
 115 K                                               & 1.2 Mil.                                   & 5.73 Mil.                                     & 12 Mil. \\ \hline
\end{tabular}}
\end{center}
\caption{Analyzed programs characteristics.}
\label{Analyzed programs statistics.}
\end{table}
\textbf{Runtime Statistics.} Table~\ref{Analyzed programs statistics.} depict several characteristics of the 2052 analyzed programs contained in the SAMATE's Juliet test suite.
In total 2052 C programs (977.7KLOC) (Juliet test suite) and 10 program (our mini-benchmark) were analyzed. The programs contained $\approx$115 K satisfiable paths, 
1.2 million unsatisfiable program execution paths and in total 12 million program branches, which 
were counted during programs analysis.

\subsection{Effectiveness (RQ1)}
\label{RQ4: Effectiveness w.r.t bug removal.}
We addressed the effectiveness of integer overflow detection by assessing the integer overflow detection rate in 2052 C programs. In order to be able to insert a
repair, the precise source code location has to be determined upfront.
First, we generated automatically 2052 Eclipse CDT compatible projects (\textit{i.e.,} for each test program a project) with the help of our tool described in~\cref{implementation}.
The Eclipse CDT projects were generated to be able to analyze the test programs with the help of the Eclipse CDT API.
Second, we generated automatically 2052 jUnit test cases (\textit{i.e.,} for each test program a jUnit test case), which were used to assess if the detected integer overflow errors were detected at the 
right source code location.
Finally, we evaluated the jUnit reports manually and cross-checked the locations of each of the detected integer overflow bug report icons. We can confirm that \textsc{IntGuard}
has achieved 100\% detection rate, \textit{i.e.,} from the total number of integer overflow errors, \textsc{IntGuard} was able to detect all. 


\subsection{False Positives and False Negatives (RQ2)}
\label{RQ2: False Positives and False Negatives}
We addressed the false positives and false negatives detection rates in order to assess the potential impact of repairing false alerts. As already mentioned, our integer overflow detection analysis is conservative, and thus our analysis is sound (\textit{i.e.,} see our soundness definition in~\cref{intro}). In other words, any detected vulnerability, which satisfies the integer overflow characteristics (\textit{i.e.,} can be confirmed by the Z3 solver) will be successfully detected and reported by \textsc{IntGuard}.

We evaluated the false positive rate of \textsc{IntGuard}, by running the previously generated jUnit test cases on all programs and assessed the rate of false positives and false negatives by checking each generated jUnit report. 
The jUnit reports confirmed by comparing the previously stored during jUnit testcase generation of the program specific information (\textit{i.e.,} line number and file name where the true positive is located) 
that each detected integer overflow was located at the expected location (\textit{i.e.,} line number and file name). Additionally, we manually checked each report in order to see if there are 
false positive or false negatives. After evaluating the results, we can confirm that we did not encounter any false positives or false negatives.

In case \textsc{IntGuard} would repair false positives then:
(1) unwanted program behavior would be inserted, 
(2) the repair would be no longer sound, 
(3) we would not be sure if integer underflows are avoided, and 
(4) by repairing a false positive the repair would be useless.

However, we are aware that by running \textsc{IntGuard} on real software we could encounter false positives due to the fact that 100\% path 
coverage is currently not achievable in practice for all types of programs. In fact, our symbolic analysis is context-sensitive, with 
high path coverage. But, nevertheless, infeasible paths due to loop unrolling may result in false positives. The conservative symbolic 
analysis could also add false positives due to possible analysis imprecision (\textit{e.g.,} due to the used environment functions). 
One possible solution is the use of concolic execution, which has the potential to further reduce imprecision due to not using environment
functions and loop invariants that are well known sources of imprecision.

\subsection{Computational Overhead (RQ3)}
\label{RQ2: Computational overhead.}
We assessed the computational overhead of generating program repairs by comparing against the runtime needed to detect 
the integer overflow error. The runtime needed by \textsc{IntGuard} to detect 2052 integer overflow errors was around 3942 
seconds (around 3 hours) and the time needed to generate the repairs was 47 seconds. This corresponds to a computational 
overhead of around 1\% (\textit{i.e.,} $3942/47 = 83.87$ seconds $\approx$1\%), which is in our opinion quite low w.r.t. the integer overflow error detection time. In addition, the repairs are useful and in many aspects superior to manually or compiler generated repairs.

\subsection{Preserving Program Behavior (RQ4)}
\label{RQ5: Preserving Program Behavior.}
We addressed whether program behavior is preserved by running \textsc{IntGuard} on all programs and by repairing all detected integer overflows.
Thereafter, we automatically compared the output log of each program against the previous log of the unrepaired program by using the same program input.
We did not observe any program execution deviation from the previous output log; just the log-line mentioning that an integer overflow bug could manifest was not present in the log of the repaired programs.

Next, we manually analyzed all source code locations where repairs were inserted and noticed that all insertion locations were 
correct locations which were previously specified in the jUnit test case. Thus, for all programs the errors were successfully removed by inserting the repair at the corresponding location.

Furthermore, after applying the repair we ran the analysis again on each of the programs in order to see if there is a potential new integer overflow error or the old one is still
present in the repaired program. We also recompiled each of the programs to investigate if the repaired program is compilable and thus syntactically correct.
We manually inspected all problem reports generated for each of the detected and repaired integer overflow bugs, and we could not detect any new integer overflow errors
for other locations in the already repaired programs. Finally, we can confirm that the behavior of the repaired programs did not change after applying the repairs.

\subsection{Benchmark Correctness Evaluation (RQ5)}
\label{Microbenchmark Correctness Evaluation}
The structural complexity of C/C++ programs makes it hard 
to assess the correctness of symbolic execution based tools against each other.
Thus, we propose a custom-build micro-benchmark to prove the 
correct detection and repair of integer overflows with our tool and any other 
future integer overflow repair tools. The micro-benchmark contains several programs
ranging from $\approx$6K to $\approx$20 LOC. Note that the number of LOC of our programs
are comparable with the LOC of popular software such as GNU Coreutils (15,065 LOC), and bzip2 (5,823 LOC). Also, each 
benchmark program contains exactly one seeded true positive integer overflow and a variable number of 
false positive integer overflows.

The goal of the micro-benchmark is to cover complex control flow scenarios including a large number of branches 
and symbolic variables, with the main objective of correctly assessing the seeded integer overflows.
We designed the micro-benchmark to cover multiple paths with variable length 
based on the following considerations. First, the total number of function 
calls inside the micro-benchmark is parameterizable. Second, the number of loop iterations 
is also parameterizable. Third, each generated program contains exactly one true positive and 
a variable number of false positives. Finally, the true positive is located deep inside the program;
several thousand of branches nested inside the program execution tree w.r.t. the root node of the tree.

Using these limits, we generated several programs and included each 
of these programs in multiple source files. These program files 
make up our micro-benchmark, which we will use to evaluate the correctness
and precision of our integer overflow detection tool.

We evaluate our micro-benchmark w.r.t. (1) time needed to run the analysis on the programs, and (2) false positive and true positive rates.
The precision of detection is evaluated by manually inspecting the reports generated by our tool 
and deciding if the report belongs to a true positive or not. The time needed to detect the true positive inside 
a program is computed by our tool and it is based on the difference between bug detection time and the start of the 
program analysis time.

\begin{table}[H]
 \begin{center}
 \resizebox{\columnwidth}{!}{%
\begin{tabular}{ |r|r|r| } 
 \hline
 \shortstack{6 KLOC Programs}       & \shortstack{11 KLOC Programs}  & \shortstack{20 KLOC Programs}   \\ \hline
 46 seconds                         & 151 seconds                    & 567 seconds \\ \hline
\end{tabular}}
\end{center}
\caption{Average analysis time in seconds over 10 runs.}
\label{Average analysis time in seconds over 10 runs.}
\end{table}
Table~\ref{Average analysis time in seconds over 10 runs.} depicts the average static analysis runtimes over 10 runs for each 
of the mini-benchmark programs grouped in three main categories based on their LOC. \textsc{IntGuard} was able to detect and repair 
all true positives present in the analyzed programs without repairing any false positives.

While most of the static analysis tools aim for detecting of a few true positives with relative high percentage of false positives
we aim to have as few false positive as possible and as many true positives as possible.

We used cppcheck (also we considered Sift which is however not open source) on our micro-benchmark, which was not able to detect any of the inserted integer overflows.
\textsc{IntGuard} was able to detect all inserted integer overflows with no false positives generated. 
Further, we found our benchmark extremely valuable during the implementation phase of our tool to detect 
corner cases of possible true positives.

As we are not aware of any currently available integer overflow tool which performs a systematic assessment of its 
repair rate on a seeded benchmark (containing large and complex programs) with integer overflows, we propose that our benchmarks or a similar 
one should be used broadly such that tools can be coherently compared against each other.
This will assure that new techniques can be easily evaluated against existing techniques and this 
further helps to push the bar towards more useful and sound tools. To faciliate these goals, we will release our micro-benchmark 
as open source.

\subsection{Runtime Performance Overhead (RQ6)}
\label{RQ3: Run-time overhead.}
We evaluated the runtime performance overhead by running \textsc{IntGuard} on all programs and by repairing all encountered integer overflow errors. We compared the runtime of the unrepaired programs with the repaired programs and noticed on average around 1\% runtime overhead.

The obtained runtime overhead is quite low since the tested programs have rather low complexity and the bugs do usually not reside in \textit{hot} code. Thus, the inserted repairs do not considerably influence the runtime overhead of the repaired program. We are aware that for more complex applications the runtime overhead can be substantially larger as a result of the program repairs, when these reside, for example, in \textit{hot} code (\textit{i.e.,} recursive calls). However, the repairs have higher priority than leaving them out, and such integer overflow errors should not be tolerated by programmers.

\subsection{Source Code and Binary Blowup (RQ7)}
\label{RQ6: Source code and binary blow-up.}
We assessed the source code and binary blowup by counting the increase in source code lines and in bytes for the resulted program binaries before and after applying the repairs.
First, we compared the total line count of the source code against the number of lines of code which were added after inserting all the repairs into the programs. As already mentioned, the 
initial line count was 977.7KLOC. After applying all repairs we added in total around 10 KLOC. This corresponds to an 
increase of less than 2\% in LOC and represents in our opinion a tolerable source code increase. Note that no code lines were deleted or compacted after applying the repair.
Second, we compared the size in bytes of the original program binary and the program binary after applying the repairs. The original size of all 2052 vulnerable programs is 1922.8 Mb.
After applying all repairs to the programs, we noticed no binary size increase. This is because we add a rather small number of checks per program. Thus, we confirm that the size of each program binary did not increase more than 1\%. This makes \textsc{InRep} highly effective and usable in, for example, embedded scenarios where minimizing program binary size is a key objective.

\subsection{Deployment (RQ8)}
\label{RQ1: Deployment.}
As one of the main design goals of \textsc{IntGuard} is automatic deployment, we describe our 
experience of applying \textsc{IntGuard} to the 2052 C programs.
\textsc{IntGuard} was able to successfully repair and recompile the 2052 C programs without any crashes.
At the same time, we were able to repair all detected integer overflow errors.

In contrast, CIntFix was not able to analyze all 2052 programs (\textit{i.e.,} only 1938 programs) since CIntFix could not deal with programs which depend heavily
on I/O (input/output) and thus the authors claim that these programs are unsuitable for automatic evaluation.
Also, CIntFix performs less informed program repairs than \textsc{IntGuard} because the repair generation and
insertion do not rely on a previous integer overflow error detection step, which usually helps to confirm that the integer overflow error is present in the program
at a particular source code line. Further, CIntFix is a runtime based tool which incurs a rather high performance overhead (\textit{i.e.,} 16\% in average) 
when tested on the same programs as \textsc{IntGuard} which has around 1\% runtime performance overhead. Finally, \textsc{CIntFix} does not provide any GUI which 
could aid the programmer to more easily locate the source code line where a repair will be inserted (aided by a report created for each integer overflow).

\subsection{Security Analysis (RQ8)}
\label{RQ7: CRAs prevention}

\begin{table}[t]
\centering 
\resizebox{.99\columnwidth}{!}{%
    \begin{tabular} {|l|l|p{1.7cm}|p{1.7cm}|c|c|c|c|} \hline
    \textbf{CVE Number} & \textbf{Application}             &\textbf{Heap Corruption}      & \textbf{Stack Corruption}    &\textbf{Local} &\textbf{Remote}   &\textbf{Patch}     &\textbf{Avoidable}        \\ \hline
           CVE-2017-797 & Ghostscript                      &                              & $\checkmark$                 &               &$\checkmark$                  &\cite{fix1}        &$\checkmark$              \\ \hline
         CVE-2016-10164 & libXpm 3.5.12                    &$\checkmark$                  &                              &$\checkmark$   &                  &\cite{fix2}        &$\checkmark$  \\ \hline
          CVE-2016-8706 & Memcached 1.4.31                 &$\checkmark$                  &                              &               &$\checkmark$      &\cite{fix3}                   &$\checkmark$  \\ \hline
          CVE-2016-9427 & bdwgc                            &$\checkmark$                  &                              &$\checkmark$   &                  &\cite{fix4}                   &$\checkmark$  \\ \hline
          CVE-2014-9862 & bsdiff, in Mac OS X              &$\checkmark$                  &                              &               &$\checkmark$      &\cite{fix5}                   &$\checkmark$  \\ \hline             
    \end{tabular}}
    \caption{Avoidable vulnerabilities by using \textsc{IntGuard}.} 
    \label{Avoidable Code Reuse Attacks.}  
\end{table}
Table~\ref{Avoidable Code Reuse Attacks.} stems from NVD~\cite{nvd} and presents a brief analysis of arbitrary code executions, 
which are caused by exploitable integer overflows. Next to the corresponding CVE number/type and the repairs used to fix the reported integer overflow vulnerability,
we highlight the fact that most of the exploitable integer overflow vulnerabilities, which lead to arbitrary code executions, are heap-based which is also confirmed in Figure~\ref{Integer overflows reported over the last 10 years.}.

We analyzed the used repairs (see Table~\ref{Avoidable Code Reuse Attacks.}) and made several observation w.r.t. the repairs:
(1) they are relatively small (up to 5 LOC),
(2) they consist of changing the wrong data type used for declaring an integer to the right data type, and
(3) they introduce some simple bounds or a cascade of bound checks on the integer value before using it 
in an unsafe program location where for example user input in form of files (or other forms of input) is used by the program as parameters. 
Another observation is that the integer overflows are not spread over a large context (only 1-2 files) and appear to have no complex dependencies, which would cause difficulties for any
static program analysis technique.

Furthermore, based on these observations we think that some of the previously depicted memory corruptions can be avoided if \textsc{IntGuard} is used consistently by
programmers in their daily routines. We make this claim based on the fact that the repairs introduced by \textsc{IntGuard} are specifically addressing the above observations, namely 
(1) the repairs check if the data type of the integer could overflow and cause an error, 
(2) the repairs are automatically adaptable in order to generate complex range checks (which can cover complex checks), and
(3) \textsc{IntGuard} can be used as a point-wise program repair insertion tool which means that it can generate
complex repairs that for humans are difficult to generate by relying on a fast local context program analysis.
Finally, we want to emphasize that \textsc{IntGuard} can also protect against other types of vulnerabilities (\textit{i.e.,} buffer overflows) 
and attacks (\textit{i.e.,} CRAs, DoS), which are based on integer overflow vulnerabilities 
by extending its integer overflow types detection capabilities.

\subsection{Controlled Experiment (RQ10)}
\label{Controlled Experiment}

\textbf{Setup.} We performed a controlled experiment by asking 30 graduate students (16 male and 14 female) 
with 1-2 years programming experience to assess \textsc{IntGuard} during a bug bounty experiment.
We split the 30 participants in two groups. The number of females and males was split evenly between the two groups.
We randomly selected three programs contained in our mini-benchmark.
We did not tell the participants which type of bug to look for and how many he/she should detect and repair.
The computer used in our experiment was equipped with two versions of Eclipse CDT (\textit{i.e.}, one Eclipse version with \textsc{IntGuard} installed and the other Eclipse 
version without \textsc{IntGuard} installed). Before the experiment was started we asked each of the participants to notify the person overlooking the experiment
when a repair was inserted and he/she has finished his/her analysis for time keeping reasons.
Next, we informed each participant from each group to find the bugs in the three given programs and repair them.

\textbf{Experiment.} 
The experiment was conducted with each participant individually placed at a single PC 
with an additional person in the room who overlooked the experiment.

\textit{Group 1.} 
Each participant had access to the most recent Eclipse CDT IDE and to the GCC (v. 4.9.3) compiler through terminal access.
Next, we asked the participants to search for bugs and fix them with the help of the Eclipse CDT IDE where \textsc{IntGuard} was \textit{not} installed. 

\textit{Group 2.}
Before the experiment, each participant from the second group got a short one minute demo movie showing them how to use \textsc{IntGuard}. Next, we asked the participants to search for bugs and fix them with the help of the Eclipse CDT IDE were \textsc{IntGuard} was installed. 

We measured the time needed for each participant to locate the bug and repair it as well as the success rate for each
analyzed program after the participant decided that he was finished. 
After the experiment, we asked each participant if (1) he/she would reuse our tool in his/her daily routine and if (2) he/she would recommend it to other peers. Each question should be answered with yes/no.

\textbf{Results.} In total, the participants needed more than 18 times (6534 vs. 362 seconds) more time to 
repair the programs without \textsc{IntGuard} than with it. From the 98 program repairs introduced manually
61\% were correct (\textit{i.e.,} the integer overflow bug was removed and no new vulnerability was introduced). 
We assessed this through manual inspection of the repairs inserted by each participant after he/she left the experiment room.
In contrast, each participant which used \textsc{IntGuard} could remove all bugs successfully. From, the total of 
30 participants 90\% (27 participants) found the tool useful and 83.3\% (25 participants) would further recommend \textsc{IntGuard} to their peers.
Overall, the results show that the time needed to find and repair an integer overflow manually is substantially higher
than with the help of \textsc{IntGuard}, and at the same time working without \textsc{IntGuard} led to a comparatively 
low rate of correct repairs.

\subsection{Comparison with Other Tools (RQ11)}
\label{Comparison with Other Tools}
First, we analyzed the 2052 C programs with cppcheck~\cite{cppcheck}. The tool reported 2592 format strings warnings (\textit{i.e.,} \%u in format string (no. 1) requires 'unsigned int' but 
the argument type is 'size\_t {aka unsigned long}), 0 errors, 1524 style issues, 1467 performance issues,
1344 portability issues (\textit{i.e.,} scanf without field width limits can crash with huge input data on some versions of libc), and 1 information 
issue (\textit{i.e.,} Cppcheck cannot find all the include files).
Second, we analyzed the 2052 C programs with the Coverity~\cite{coverity} open source static analysis tool. The reports generated for the analyzed programs indicated several 
possible integer overflows, but \textit{no} integer overflow repairs were suggested.
Third, we compiled all programs, which we analyzed, with all warnings that GCC (v. 4.9.3) has to offer.
We parsed all the logs generated and observed that neither any integer overflow reports nor source code repairs were suggested.
Fourth, we compiled all 2052 programs with all warning flags on with Clang (v. 3.8.0). We parsed all log outputs
and observed no integer overflows repair suggestions. Even though the warnings for Clang are more expressive than those of GCC, we did not observe any suggestion on how to repair the compiled programs.
In summary, we can confirm that all tools, which we used in our evaluation, are not able to suggest any code repairs that can be used to avoid the integer overflows present in the analyzed programs.

\section{Discussion}
\label{discussion}
In this section, we compare in \cref{Comparison with other Tools} \textsc{IntGuard} with 
other runtime tools, 
and in \cref{Limitations} we present some limitations of \textsc{IntGuard}.

\subsection{Comparison with other Tools}
\label{Comparison with other Tools}

\subsubsection{Source Code Based Tools}
CIntFix~\cite{CIntFix} is a runtime based tool used for repairing of integer related problems in C source code functions. 
This tool utilizes integers of infinite size with two’s complement encoding in place of
original bounded integers. However, CIntFix cannot automatically infer constraints on arguments
of library function without previously provided code annotations. CIntFix is
unable to tolerate intentional wraparounds which directly
propagate to critical sites. 
Thus, CIntFix was not able to analyze all 2052 programs since 
CIntFix can not deal with programs which depend heavily
on I/O. Furthermore, CIntFix does not rely on integer overflow repair as \textsc{IntGuard} does, thus it na{\"i}vely inserts repairs at all possible
source code locations, even at the ones where no integer overflow can occur at all. In contrast, \textsc{IntGuard} has $\approx$ 1\% overhead compared to
CIntFix (\textit{i.e.,} 16\% is low). CIntFix has a high source code expansion of 25\% while \textsc{IntGuard} has $\approx$ 2\% source code expansion due to
the fact that the repairs inserted by \textsc{IntGuard} are inserted more informed.
Finally, CIntFix does not have any GUI support during repair insertion which can be a considerable advantage in some situations.

\subsubsection{Program Binary Based Tools}
TAP~\cite{tap} is a runtime based tool which operates directly on x86 binaries. 
It is the most similar tool to \textsc{IntGuard} since both tools 
first detect an integer overflow and next they generate a code repair which removes the integer overflow. 
TAP's integer overflow discovery algorithm is based on Diode~\cite{targeted}. 
TAP needs access to the program's source code in order to insert the generated repair.
TAP monitors the execution of the application to identify memory allocation sites and construct symbolic
expressions that capture the size of the allocated buffer as a function
of the input bytes. It then uses goal-directed conditional branch
enforcement to generate inputs that (1) overflow the computation of
the size of the allocated buffer while (2) forcing the application to take
a path that executes the statement at the memory allocation site. Repair generation is based on
templates that are matched against a previously generated symbolic expressions. 
If a template matches, TAP applies the template to generate an associated patch and inserts the patch into
the application. Compared to \textsc{IntGuard}, TAP relies on seed inputs which if not available limit the usability of TAP.
Also, TAP checks that the integer overflow error was removed with a limited set of input seeds which do not guarantee 
that the error was really removed from the 
program binary. Finally, we note that the number of repair guarantees offered by TAP fulfill less guarantees than and thus are less 
safe as those offered by \textsc{IntGuard}.

\subsection{Limitations}
\label{Limitations}
First, our evaluation is based on the currently largest open source test suite for C/C++ programs. Thus,
the findings of this evaluation still do not necessarily reflect the behavior of \textsc{IntGuard} when applied to larger programs.
However, we think that this does not limit the applicability of \textsc{IntGuard} since our tool is highly scalable 
due to its configurable analysis and the potential for implementation of fuzzing techniques which make our tool even more effective.

Second, the implementation of \textsc{IntGuard} depends on loop unrolling which incurs well known precision penalties. 
This insufficiency can be addressed in some cases by a prior analysis of program loops in order to derive loop invariants 
and in cases where this is not statically determinable symbolic analysis could switch to concolic analysis.
Our static analysis is time-consuming.  Its accuracy and performance will affect \textsc{IntGuard} results. 
Furthermore, environment functions provided by the programmers are needed.

Third, another potential limitation for certain programers is the fact that \textsc{IntGuard} is developed 
as a plug-in and runs inside Eclipse CDT. This could be annoying to some programmers used to command line tools. 
Nevertheless, this can be easily addressed by exporting \textsc{IntGuard} as a command line tool similar to
CIntFix~\cite{CIntFix} which relies as \textsc{IntGuard} on the Eclipse CDT framework as well.

Fourth, at this stage of development our static analysis engine can build the control flow graph of the program and analyze a subset of the C/C++ programming languages.
This means that only certain types of statement and functions headers can be understood by our engine. This limitation can be eliminated by working the complete 
list of C/C++ statement types and function headers. This can be solved in future by investing sufficient time and manpower.

Finally, we tested \textsc{IntGuard} in a controlled experiment with a restricted number of participants and for this reason our
findings will do not necessarily scale in industrial settings were real development conditions are available.
Nevertheless, we think that our tool can help to drastically cut down the time needed for bug finding an repair due to its usability
and low intrusiveness.

\section{Related Work}
\label{related}
Integer overflows have threatened software programs for decades.
The usage of different types of tools based on:
{static analysis integer overflow detection}~\cite{sift, targeted, kint},
{runtime program repair}~\cite{diehard},
{benign integer overflow identification}~\cite{kint}, 
{directed and random fuzzing}~\cite{buzzfuzz, taintscope},
{concolic testing}~\cite{klee}, 
{library support and runtime checks}~\cite{integerlib, safeint}, and
{repair code transfer}~\cite{codephage} have helped to reduce the number of integer overflows.
These tools have seen only little to no adoption in the industry; partly because their benefits are
hard to assess in the context of real software projects~\cite{benefits} where there is an urgent need  
to detect and avoid integer overflow based memory corruptions,
which can lead to CRAs~\cite{rop:buchanan} or
other security vulnerabilities~\cite{wang:intscope, kint}.

\begin{table}[t]
\fontsize{11}{9}
\begin{threeparttable}
\centering
\noindent\makebox[\linewidth]{%
\resizebox{\linewidth}{!}{%
{\LARGE %
  \begin{tabular}{|l|l|c|c|c"c|c"c|c|c"c|c"c|c|c|c"c|c"c|c"c|c|c"c|c|c|c|}\hline 
      \rotatebox{0}{Tool}                  
    & \rotatebox{90}{s. s. d.}             
    & \rotatebox{90}{s. $\neg$ s. d.}      
    & \rotatebox{90}{d. s. d.}             
    & \rotatebox{90}{d. $\neg$ s. d.}      
    & \rotatebox{90}{SMT solv.}            
    & \rotatebox{90}{$\neg$ SMT solv.}     
    & \rotatebox{90}{source code}          
    & \rotatebox{90}{binary code}          
    & \rotatebox{90}{intermediate}         
    & \rotatebox{90}{C support}            
    & \rotatebox{90}{C++ support}          
    & \rotatebox{90}{overflow}             
    & \rotatebox{90}{underflow}            
    & \rotatebox{90}{signedness}           
    & \rotatebox{90}{truncation}           
    & \rotatebox{90}{sound}                
    & \rotatebox{90}{complete}             
    & \rotatebox{90}{benign}               
    & \rotatebox{90}{exploitable}          
    & \rotatebox{90}{annotations}          
    & \rotatebox{90}{fuzzing}              
    & \rotatebox{90}{mutating}             
    & \rotatebox{90}{detect (d)}        
    & \rotatebox{90}{repair (r)}           
    & \rotatebox{90}{(d\&r)} \\ \hline     
    \hline
    ARCHER~\cite{archer:esorics}           &  &  &  &\c&  &\c&\c&  &  &\c&  &\c&\c&  &  &  &  &  &  &\c&  &  &\c&  &\\ \hline
    UQBTng~\cite{uqbtng:wojtczuk}          &  &  &  &\c&  &\c&  &\c&  &  &  &\c&  &  &  &  &  &  &\c&\c&  &  &\c&  &\\ \hline
    PREfast~\cite{prefast:microsoft}       &  &\c&  &  &  &\c&\c&  &  &\c&\c&\c&  &  &  &  &  &  &  &\c&  &  &\c&  &\\ \hline
    Rich~\cite{brumley:rich}               &  &  &  &\c&  &\c&\c&  &  &\c&  &\c&\c&\c&\c&  &  &\c&\c&  &  &  &\c&  &\\ \hline
    SAGE~\cite{godefroid:sage}             &  &  &\c&  &\c&  &  &\c&  &  &  &\c&\c&\c&\c&  &  &  &  &  &\c&  &\c&  &\\ \hline
    IntScope\cite{wang:intscope}           &\c&  &  &  &\c&  &  &  &\c&  &  &\c&  &  &  &  &  &  &  &  &  &\c&\c&  &\\ \hline
    Brick~\cite{chen:brick}                &  &  &  &\c&  &\c&  &  &\c&  &  &\c&\c&\c&\c&  &  &  &  &  &  &  &\c&  &\\ \hline
    IntFinder~\cite{intfinder}             &  &  &  &\c&  &\c&  &\c&  &  &  &\c&\c&\c&\c&  &  &  &  &  &  &  &\c&  &\\ \hline 
    SmartFuzz~\cite{molnar:smartfuzz}      &  &  &\c&  &\c&\c&  &  &\c&  &  &\c&\c&\c&\c&  &  &  &  &  &\c&  &\c&  &\\ \hline
    PREfix~\cite{moy:prefix}               &\c&  &  &  &\c&\c&  &  &  &\c&\c&\c&  &  &  &\c&  &  &  &  &  &  &\c&  &\\ \hline
    IntPatch~\cite{int:patch}              &  &\c&  &  &  &  &\c&  &  &\c&\c&\c&  &  &  &\c  &  &  &  &  &  &  &\c&  &\\ \hline
    IOC~\cite{dietz:ioc}                   &  &  &  &\c&  &\c&\c&  &  &\c&\c&\c&\c&\c&\c&  &  &  &  &  &  &  &\c&  &\\ \hline
    IntFlow~\cite{intflow:hybrid}          &  &  &  &\c&  &\c&\c&  &  &\c&\c&\c&\c&\c&\c&  &  &  &  &  &  &  &\c&  &\\ \hline
    SoupInt~\cite{soupint}                 &  &  &\c&  &\c&  &  &\c&  &  &  &\c&  &  &  &  &  &\c&\c&  &  &  &  &\c&\\ \hline
    SIFT~\cite{sift}                       &  &\c&  &  &  &\c&\c&  &  &\c&  &\c&  &  &  &\c&  &  &  &\c&  &  &  &\c&\\ \hline 
    TAP~\cite{tap}                         &\c&  &  &  &\c&  &\c&\c&  &\c&\c&\c&  &  &  &  &  &  &\c&  &  &  &\c&\c&\c\\ \hline
    Diode~\cite{targeted}                  &\c&  &  &  &\c&  &  &\c&  &  &  &\c&  &  &  &  &  &  &  &  &  &  &\c&  &\\ \hline
    Indio~\cite{integer:binary}            &\c&  &  &  &\c&  &  &\c&  &  &  &\c&  &  &  &  &  &  &\c&  &  &  &\c&  &\\ \hline
    Zhang et al.~\cite{toolx}              &  &  &\c&  &\c&  &  &\c&  &  &  &\c&  &  &  &  &  &  &  &  &  &  &\c&  &\\ \hline 
    IntEQ~\cite{inteq}                     &\c&  &  &  &\c&  &\c&  &  &\c&\c&\c&\c&\c&\c&  &  &\c&\c&  &  &  &\c&  &\\ \hline
    CIntFix~\cite{CIntFix}                 &  &  &  &\c&  &\c&\c&  &  &\c&  &\c&\c&\c&\c&  &  &  &  &\c&  &  &  &\c&\\ \hline \hline
     
    \hline 
    IntGuard                               &\c&  &  &  & &  &\c  &  &  &\c&  &\c&\c&  &  &\c &  &  &  &  &  &  &\c&\c&\c \\ \hline
  \end{tabular}}}}
  \caption{Integer overflow detection and repair features.}
\label{tab:truthTables:bla}   
\end{threeparttable}
\end{table} 
Table~\ref{tab:truthTables:bla} summarizes 
a string of tools used for integer overflow detection and repair, and the underlying techniques on which these tools are based. 
In Table~\ref{tab:truthTables:bla} the \textbullet \ symbol means addressed and the abbreviations have the following meaning: static symbolic detection (s. s. d.), 
dynamic symbolic detection (d. s. d), and $\neg$ logical not.
We deliberately excluded from Table~\ref{tab:truthTables:bla} commercial static symbolic execution tools (\textit{e.g.,} CodeSonar; see~\cite{commercial:tools} for more details), 
which scale to programs having millions LOC and which can efficiently detect integer overflows, since their internals are mostly unknown. 

It is interesting to note that most of these tools are either only used for integer overflow detection or if they are used for code repair then they do not 
consider detecting first the integer overflow. More specifically, \textsc{IntGuard} and TAP are the \textit{only} tools which first detect the integer overflow 
and than propose a repair for it. Additionally, most of the other repair generation tools are repairing the programs in an uninformed (\textit{i.e.,} without 
first detecting and classifying the bug) manner, which consequently incurs high runtime overhead and significant insertion imprecision. Further, only SIFT is 
based on a sound technique, and none of the other tools are complete for well known reasons---discussing all these reasons here is out of scope of this paper. 
In contrast, \textsc{IntGuard} code repairs are sound (\textit{e.g.,} no false positives, see our soundness definition in~\cref{intro}). Please note that comparing
all the tools depicted in Table~\ref{tab:truthTables:bla} against~\textsc{IntGuard} in detail is not feasible within the page length constraints of this paper.
Thus, next we briefly mention several integer overflow repair generation tools and relate them to \textsc{IntGuard}.

\subsubsection{Source Code Repairs}
CIntFix~\cite{CIntFix} is a runtime based tool used for protection against integer related problems in C source code functions only.
This tool utilizes integers of infinite size with two's complement encoding in place of original bounded integers. 
The code transformations of CIntFix can be only applied to complete functions and the analysis is not inter-procedural. Further,
the runtime slowdown is around 18\%. CIntFix has several advantages. The analysis performed by CIntFix is 
syntax-directed and rule-based, which avoids sophisticated and imprecise analysis. Its analysis cannot scale to large programs and
CIntFix can miss the repair of some integer overflows as well. CIntFix is unable to tolerate intentional wraparounds, which directly propagate to critical sites.
From the total of 2052 programs in CWE190, it could repair only 1938 (114 misses), whereas \textsc{IntGuard} could fix all of them. Also, the source code expansion
is above 25\%, which is $\approx$25 times more than the code expansion of \textsc{IntGuard}. The runtime on the repaired programs is 16\%, whereas \textsc{IntGuard}
slowdown on the same programs is $\approx$1\%. CIntFix claims to provide transformation, which are safe, but this is currently hard to assess since its transformations
and the ones of AIC/CIT/RAO~\cite{coker:integers} are very similar and the transformations of the later do not preserve program behavior.

AIC/CIT/RAO~\cite{coker:integers} is a static source code analysis tool built as an Eclipse plug-in, which provides several code transformations that can be used to avoid integer overflows. The provided transformations are similar to code refactorings, but actually transform the internal model of a C program 
towards a safe model. This tool provides three types of transformations: 
add integer cast (AIC), replace arithmetic operator (RAO), and change integer type (CIT). Based on these three safe code transformations, this tool 
can protect against integer overflows without the need to detect the integer overflow first. In contrast, \textsc{IntGuard} first detects the integer overflow
using symbolic static analysis and then it generates a repair, which can protect against the integer overflow and underflow.
The above three transformations can not be applied to all unsafe situations and a fraction of the variable declarations are also modified since in some
situations the preconditions which have to be checked are far to complex than what AIC/CIT/RAO can cover. In this situation no transformations are applied.
The tool has an runtime overhead of the hardened programs which is over 30\% which in real software is not acceptable.
The AIC/CIT/RAO transformations do not preserve the original program behavior as mentioned in the original paper. The goal of these transformations is to transform the program to a more 
safe integer model. The CIT transformation suffers from the need of justification
of a type change. Finally, the repairs cannot be considered user-friendly resulting often in complex cascaded transformations, which are hard to assess by a programmer.

\subsubsection{Binary-Based Repairs}
Sift~\cite{sift} is a binary based which can protect programs against integer overflows by instrumenting its binary.
Sift relies on carefully crafted user source code annotations in order to identify the input field that each input statement reads.
The tool can not generate input filters for all types of source code sites (\textit{i.e.,} currently only memory allocations 
and block memory copy sites). In contrast, \textsc{IntGuard} does not rely on source code annotations.
Note that the integer overflow triggering file input formats are all image based (\textit{i.e.,} PNG, JPEG, GIF, SWF) or sound (\textit{i.e.,} WAV) based.
Particularly, when expressions at a site contain subexpressions whose values depend on an unbounded number of values computed in loops the tool needs an 
upper bound on the number of loop iterations which can be specified by the programer upfront (\textit{i.e.,} currently not used by the tool).
\textsc{IntGuard} takes a different approach by statically unrolling each loop 10 times.
Sift produced no false positives by applying 62K inputs to the previous repaired programs.
These still does not mean that the tool cannot produce false positives. In order to filter out an integer overflow a potentially infinite input
set has to be applied to the program in order to be 100\% sure that there is no false positive. 
In contrast, \textsc{IntGuard} does not rely on any input integer overflow triggering files which arguably in for some types of programs are much more harder to be generated by Sift 
(\textit{i.e.,} configuration files for web servers, etc.).
Further, Sift may introduce unwanted program behavior since it does not first detect the integer overflow but rather it 
generates an input filter for the previously annotated integer overflow prone source code site. In contrast \textsc{IntGuard} detects first the integer
overflow and than it proposes a repair for it which further is not na{\"i}vely inserted since the final insertion decision is delegated to the programmer.
TAP~\cite{tap} is yet another runtime based tool used for repairing C programs by operating on x86 program binaries. TAP
is similar to \textsc{IntGuard} w.r.t. the fact that both tools first detect an integer overflow and next they generate a code repair which 
removes the integer overflow. TAP utilizes the integer overflow discovery algorithm from Diode~\cite{targeted}
and needs access to both source code and the binary in order to insert the generated repairs in the program binary.
It checks that the integer overflow error was removed with a limited set of input seeds which do not guarantee that the error 
was really removed from the program binary since the set of provided inputs is limited.

\subsubsection{Transferring Code Repairs}
This techniques assume that several applications which run on multiple input files
and an input file which triggers an error in one application can be used (for example CodePhage~\cite{codephage}) to 
find checks in other applications that enable other applications to successfully process the input file.
Next, it relies on multi-application code transfer to transfer these checks into the original application 
and eliminate the error. 
\textsc{IntGuard} differs in that its code repair technique enable it to generate repairs in the absence of 
any applications that need to process the same inputs and the checks do not have to be somewhere else present in the application but
rather these are completely automatically generated.

\subsubsection{Static Integer Errors Detection}
Several static analysis tools have been proposed to address integer related problems.
Diode~\cite{targeted} relies on targeted site identification and goal-directed conditional branch enforcement
in order to discover integer overflow errors in x86 binaries. Compared to our tool Diode does not exercise all bugs due to a limited 
number of test inputs.
Similarly, SIFT~\cite{sift} uses a sound static program analysis in order to generate filters that remove
inputs that may trigger overflow errors and it is not intended to be used for integer overflow error identification.
In contrast, \textsc{IntGuard} does not rely on carefully crafted inputs and is intended to be used for integer overflow error detection and repair.
KINT requires optionally procedure specifications from the programmer in order to characterize parameter value ranges
and it reports many false positives~\cite{kint}. In contrast, \textsc{IntGuard} analysis proves the existence of 
integer overflow errors without any false negatives on the evaluated programs.

\subsubsection{Benign Integer Overflows} 
Sometimes code contains benign IOs~\cite{kint}. 
A possible concern is that the integer overflow repair tools may interfere with the behavior of such programs.
Because \textsc{IntGuard} focuses primarily on critical assignments sites (\textit{i.e.,} other types can be also supported) that are mostly unlikely to
contain such intentional IOs, it is unlikely to remove benign IOs and therefore to interfere with the intended program behavior.

\subsubsection{Directed and Random Fuzzing}
Taking a different approach, fuzzing-based software testing is used by large companies, but suffers from well known limitations including a 
lack of usability of the randomly generated seed input set and the fact that the code repair is still left as a tedious and error-prone manual 
task for the programmer. An example is Google's OSS-Fuzz~\cite{oss:fuzz},
which uses this technique for bug detection, but does not provide repairs.
Directed fuzzing~\cite{taintscope, buzzfuzz} main goal is to expose errors which reside deep inside programs.
BuzzFuzz~\cite{buzzfuzz} and TaintScope~\cite{taintscope} use taint tracking to identify input bytes that influence 
values at critical program sites such as memory
allocation sites and system calls. These tools are successful at reducing the size of the fuzzed input
but in general are inefficient at finding carefully crafted inputs used to expose integer overflow errors. Because these
directed fuzzing 
tools operate on raw binary input, the changes in the input can produce syntactically incorrect 
input that fail the sanity checks. 
Another technique is based on random fuzzing which is successfully used by 
security researchers~\cite{peachfuzzer}. Due to the fact that its generated inputs fail input checks, random 
fuzzing is relatively ineffective for example at generating inputs that trigger integer overflow errors.

\subsubsection{Concolic Testing}
Concolic testing is a newer alternative than directed and random fuzzing~\cite{klee}. 
These tools execute programs both concretely and symbolically on a seed input until an interesting program
expression is reached. Although successful in many scenarios~\cite{klee}, concolic testing faces several challenges~\cite{monirul}.
Specifically, the resulted constraint systems for deeper program paths get very complex and thus beyond the 
capabilities of current SMT solvers.
SmartFuzz~\cite{molnar:smartfuzz} is a concolic testing tool which can detect IO, 
non-value-preserving width conversions, and potentially dangerous signed/unsigned conversions. 
Furthermore it is limited by deep program paths and blocking checks.
Dowser~\cite{dowser} is a fuzzer that combines taint tracking, program analysis, and symbolic execution 
to find buffer overflows.
These tools are optimized for path coverage and therefore it is unlikely to discover integer overflow errors.
TAP uses the algorithms of DIODE~\cite{targeted} to detect integer overflow errors. 
DIODE compared to other similar 
tools is targeted. It starts with a critical site that is executed by a seed input. A range of
techniques are used in order to navigate sanity and blocking checks to trigger an overflow at 
the critical site.

\subsubsection{Library Support and Runtime Checks}
Safe integer libraries such as, SafeInt~\cite{safeint}
or IntegerLib~\cite{integerlib}, are widely used during runtime. These approaches impose on the programmers
that they rewrite existing code to use 
safe integer operations. \textsc{IntGuard}, in contrast, detects and repairs integer overflow errors without any assistance 
from developers to rewrite code.
Another wide used approach to address the problem of false positives is
based on runtime detection tools that dynamically insert runtime checks before integer related 
operations~\cite{brumley:rich}. One major drawback of the inserted checks is that these incur often time a high runtime overhead. 
In contrast, \textsc{IntGuard} inserts checks only when it previously found an IO. It therefore 
imposes a low performance overhead. 
Other tools such as IOC~\cite{dietz:ioc}, IntPatch~\cite{int:patch} and 
IntEQ~\cite{inteq} add runtime checks statically by using compiler instrumentation in order to check for integer overflow errors 
during runtime when more context is available. Input rectification~\cite{rectification} is another approach 
which modifies
program inputs that crash an application such that it will not crash afterwards. Because it learns needed constraints
that the input has to satisfy these technique is susceptible to false positives.

\section{Future Work}
\label{future}
In this section we mention briefly several avenues for improving \textsc{IntGuard} in future work.
\subsection{Guided Program Path Exploration}  
For example all kind of guided 
symbolic execution techniques help to guide the symbolic analysis to more \textit{interesting} program locations by 
skyping those program locations which are less prone to contain integer overflow errors. Such techniques have been widely used in the past with 
more or less success. One typical characteristic from which \textsc{IntRep} could also benefit is the discovery of \textit{deep} integer overflow errors which usually could 
not be exercised since there are located deep in the program. 

\subsection{Program Path Pruning Techniques} 
Further, we want to implement several path pruning techniques in order to 
run \textsc{IntRep} on larger programs. These techniques can be beneficial to \textsc{IntRep} in many ways. First, 
several light-weight path exploration techniques such as DFS, BFS can be implemented in order to better quide the analysis. Second, we want to combine the previous mentioned path exploration techniques 
with path merging techniques based on dead variables or interpolation which further help to reduce the search space. We are aware that those techniques
will not overcome the path explosion problem and as such full path coverage is rather a \textit{mythical} thing to attain. On the other hand, we 
strongly believe that the main benefit of this research paths would be the detection of previously unknown integer overflow errors which would make the effort worthwhile.

\subsection{Other Reparable Error Types} 
All types (\textit{e.g.,} signedness, etc.) of integer related errors, which CIntFix can repair, can be also 
detected and repaired by \textsc{IntRep} in principle.
The main step towards this goal is to extend the set of possible checked locations for integer related problems.
Furthermore, we want to address the repair of other types of integer related problems 
(\textit{e.g.,} underflow, signedness, truncation, intentional, unintentional and undefined)
which can lead to CRAs as well. 

\subsection{Extending to C++ programs} Currently our program statement translation component is under construction and we plan in future to be able to
scale to the C++ language as well in order to be able to cover all possible language semantics. Additionally, we think that this is just a matter of time and manpower which 
has to be invested in order to achieve this goal.

\subsection{Caching Techniques} 
We want to explore how to efficiently run the integer overflow detection and repair tool in online
mode such that program related information can be cached and reused for repair generation when 
the program number of code lines increases over time. This can be achieved efficiently by using the 
information which version control systems provide.

\subsection{Industry Acceptance} 
In the future, we want to test \textsc{IntRep} in real industrial scenarios.
Most of the successfully used tools for error detection in the industry (\textit{i.e.,} at Google) are mainly based 
on fuzzers. These tools have proven to be effective and scalable to large code bases and their success rate depends heavily on the 
search heuristic behind the technique. Moreover, we think that static analysis tools should be more used by programmers and such 
tools should be particularly tailored (\textit{i.e.,} integrated with build and/or version control tools) for their 
needs in order to find acceptance. Thus, there is a still long way to go but we are confident that these tools will find wide acceptance and 
replace fuzzing only based techniques since their accuracy and effectiveness is superior to the previous mentioned tools.
%

\section{Conclusion}
\label{conclusion}
In this paper, we presented \textsc{IntGuard}, an integer overflow detection and repair tool for C source code, which provides sound, highly useful and high-quality code repairs that satisfy more 
repair guarantees than other state-of-the-art tools.

In our evaluation, we applied \textsc{IntGuard} to C programs having $\approx$1 Mil. LOC.
Experimental results show that \textsc{IntGuard's} repairs are
more effective (\textit{i.e.,} repaired programs have around 1\% runtime overhead), 
precise (\textit{i.e.,} no false positives are repaired), and 
more useful (\textit{i.e.,} sound repairs) than other state-of-the-art tools. 
\textsc{IntGuard} was able to repair integer overflows using automatically generated source code repairs that incurred less than 2\% increase in LOC and around 1\% program binary blow-up.
We conducted a controlled experiment with 30 participants in which we showed that \textsc{IntGuard} is 18 times more time-effective 
and has a higher repair success rate than manual repairs. At the same time, 91\% of the participants found \textsc{IntGuard} highly usable and 83\% of the participants
would further recommend it to their peers. Finally, we point out that in order to protect
against CRAs and other types of integer overflow based vulnerabilities a promising approach is to check the program source code 
in a consistent fashion as part of programmers' daily routine; \textsc{IntGuard} can address this goal in an efficient and programmer-friendly fashion.



\bibliographystyle{IEEEtran} 

\begin{thebibliography}{50}

\bibitem{barret:smtlib}
C. Barrett, A. Stump, and C. Tinelli,
\newblock{The SMT-LIB Standard Version 2.0},  
\path{http://smtlib.cs.uiowa.edu/papers/smt-lib-reference-v2.0-r10.12.21.pdf}, 2010.

\bibitem{diehard}
E. D. Berger, and B. G. Zorn,
\newblock {Diehard:Probabilistic memory safety for unsafe languages},
In \textit {Proceedings of the International Conference on Software Engineering (ICSE)}, 2012.

\bibitem{brumley:rich}
D. Brumley, T. Chiueh, and R. Johnson,
\newblock {RICH: Automatically protecting Against Integer-based Vulnerabilities},
In \textit{Proceedings of the Network and Distributed System Security Symposium (NDSS)}, 2007.

\bibitem{rop:buchanan}
E. Buchanan, R. Roemer, H. Shacham, and S. Savage,
\newblock {When Good Instructions Go Bad: Generalizing Return-Oriented Programming to RISC},
In \textit {Proceedings of the ACM Conference on Computer and Communications Security (CCS)}, 2008.

\bibitem{vcc}
H. Perl, 
S. Dechand, 
M. Smith, 
D. Arp,
F. Yamaguchi,
K. Rieck, 
S. Fahl, and
Y. Acar,
\newblock {VCCFinder: Finding Potential Vulnerabilities in
Open-Source Projects to Assist Code Audits},
In \textit {Proceedings of the ACM Conference on Computer and Communications Security (CCS)}, 2008.

\bibitem{oss:fuzz}
Google,
\newblock{OSS-Fuzz - Continuous Fuzzing for Open Source Software},  
\path{https://github.com/google/oss-fuzz}, 2016.

\bibitem{klee}
C. Cadar, D. Dunbar, and D. Engler,
\newblock{KLEE: Unassisted and automatic generation of high-coverage tests for complex systems programs},
In \textit {Proceedings of the Annual Computer Security Applications Conference (ACSAC)}, 2014.

\bibitem{chen:brick}
P. Chen, Y. Wang, and Z. Xin,
\newblock {Brick: A Binary Tool for Run-time Detecting and Locating Integer-based Vulnerability},
In \textit {Proceedings of the International Conference on Availability, Reliability and Security (ARES)}, 2009.

\bibitem{intfinder}
P. Chen, H. Han, Y. Wang, X. Shen, X. Yin, B. Mao, and L. Xie,
\newblock {IntFinder: Automatically Detecting Integer Bugs in x86 Binary Program},
In \textit {Proceedings of the International conference on Information and Communications Security (ICICS)}, 2009.

\bibitem{archer:esorics}
R. Chinchani, A. Iyer, B. Jayaraman, and S. Upadhyaya,
\newblock {ARCHERR: Runtime Environment Driven Program Safety},
In \textit {Proceedings of the European Symposium on Research in Computer Security (ESORICS)}, 2004.

\bibitem{CIntFix}
X. Cheng, M. Zhou, X. Song, M. Gu, and J. Sun,
\newblock {Automatic Fix for C Integer Errors by Precision Improvement},
In \textit {Proceedings of the Annual Computer Software and Applications Conference (COMPSAC)}, 2016.


\bibitem{cwe190:overflow}
Integer Overflow or Wraparound,  
\path{https://cwe.mitre.org/data/definitions/190.html}.

\bibitem{cwe191:underflow}
Integer Underflow (Wrap or Wraparound), 
\path{https://cwe.mitre.org/data/definitions/191.html}.

\bibitem{cwe192:coercion}
Integer Coercion Error, 
\newblock {\path{https://cwe.mitre.org/data/definitions/192.html}}. 

\bibitem{cwe193:offbyone}
Off-by-one Error, 
\newblock {\path{http://cwe.mitre.org/data/definitions/193.html}}. 

\bibitem{cwe194:sign}
Unexpected Sign Extension, 
\newblock {\path{https://cwe.mitre.org/data/definitions/194.html}}. 

\bibitem{cwe195:signed}
Signed to Unsigned Conversion Error,
\path{https://cwe.mitre.org/data/definitions/195.html}.

\bibitem{cwe196:unsigned}
Unsigned to Signed Conversion Error,
\newblock {\path{https://cwe.mitre.org/data/definitions/196.html}}.

\bibitem{cwe197:truncation}
Numeric Truncation Error
\path{https://cwe.mitre.org/data/definitions/197.html}.

\bibitem{commercial:tools}
S. Shiraishi, V. Mohan, and H. Marimuthu,
\newblock{Quantitative Evaluation of Static Analysis Tools},
In \textit {Proceedings of the Network and Distributed System Security Symposium (NDSS)}, 2008.

\bibitem{cwe680}
Integer Overflow to Buffer Overflow,
\newblock {\path{https://cwe.mitre.org/data/slices/680.html}}.

\bibitem{z3:smt}
L. de Moura and N. Bjørner,
\newblock {Z3: an efficient SMT solver},
In \textit {Proceedings of the International Conference on Tools and Algorithms for the Construction and Analysis of Systems/European
Joint Conference on Theory \& Practice of Software (TACAS/ETAPS)}, 2008.

\bibitem{dietz:ioc}
W. Dietz, P. Li, J. Regehr, and V. Adve,
\newblock {Understanding Integer Overflow in C/C++},
In \textit {Proceedings of the International Conference on Software Engineering (ICSE)}, 2012.

\bibitem{gcc:discussion}
Discussion between programmers and gcc developers,
 \path{http://gcc.gnu.org/bugzilla/show_bug.cgi?id=30475#c2}.

\bibitem{cppcheck}
A tool for static C/C++ code analysis,
 \path{http://cppcheck.sourceforge.net/}.

\bibitem{demo1}
Integer overflow detection demo,
 \path{https://goo.gl/uNvdRp}.

\bibitem{demo2}
Integer overflow repair demo,
 \path{https://goo.gl/912Jux}.

\bibitem{gamma:patterns}
E. Gamma, J. Vlissides, R. Johnson and R. Helm,
\newblock{Design Patterns. Elements of Reusable Object-Oriented Software},
\newblock {Addison-Wesley '94}.

\bibitem{godefroid:sage}
P. Godefroid, M. Y. Levin and D. Molnar,
\newblock{Automated Whitebox Fuzz Testing},
In \textit {Proceedings of the Network and Distributed System Security Symposium (NDSS)}, 2008.

\bibitem{buzzfuzz}
V. Ganesh, T. Leek, and M. Rinard.
\newblock{Taint-based directed whitebox fuzzing},
In \textit {Proceedings of the International Conference on Software Engineering (ICSE)}, 2009.

\bibitem{dowser}
I. Haller, A. Slowinska, M. Neugschwandtner, and H. Bos,
\newblock {Dowsing for overflows: a guided fuzzer to find buffer boundary violations},
In \textit {Proceedings of the USENIX Security Symposium (USENIX SEC)}, 2013.

\bibitem{codan:laskavaia}
E. Laskavaia,
\newblock{Codan: a C/C++ Static Analysis Framework for CDT},  
In \textit {EclipseCon '11}.

\bibitem{pldi:le}
V. Le, M. Afshari, and Z. Su,
\newblock{Compiler Validation via Equivalence Modulo Inputs},
In \textit {Proceedings of the Annual Computer Security Applications Conference (ACSAC)}, 2014.

\bibitem{rectification}
F. Long, V. Ganesh, M. Carbin, S. Sidiroglou, and M. Rinard.
\newblock{Automatic input rectification},  
In \textit {Proceedings of the International Conference on Software Engineering (ICSE)}, 2012.

\bibitem{sift}
F. Long, S. Sidiroglou-Douskos, D. Kim, and M. Rinard,
\newblock {Sound Input Filter Generation for Integer Overflow Errors},
In \textit {Proceedings of the Symposium on Principles of Programming Languages (POPL)}, 2014.

\bibitem{prefast:microsoft}
Microsoft,
\newblock{PREfast analysis tool},  
 \path{https://msdn.microsoft.com/en-us/library/ms933794.aspx}, Microsoft Corporation, 2006.

\bibitem{molnar:smartfuzz}
D. Molnar, X. C. Li, and D. A. Wagner,
\newblock{Dynamic Test Generation to Find Integer Bugs in x86 Binary Linux Programs},
In \textit {Proceedings of the USENIX Security Symposium (USENIX SEC)}, 2009.

\bibitem{moy:prefix}
Y. Moy, N. Bjørner, and D. Sielaff,
\newblock{Modular Bug-finding for Integer Overflows in the Large: Sound, Efficient, Bit-precise Static Analysis},
\newblock {MSR-TR-2009-57}, 2009.

\bibitem{muntean:integer}
P. Muntean, M. Rahman, A. Ibing and C. Eckert,
\newblock{SMT-Constrained Symbolic Execution Engine for Integer Overflow Detection in C Code},
In \textit {Proceedings of the International Information Security South Africa Conference (ISSA)}, 2015.

\bibitem{muntean:buffer}
P. Muntean, V. Kommanapalli, A. Ibing and C. Eckert,
\newblock{Automated Generation of Buffer Overflows Quick Fixes using Symbolic Execution and SMT},
In \textit {International Conference on Computer Safety, Reliability \& Security (SAFECOMP)}, 2015.

\bibitem{juliet:test}
U.S. National Institute of Standards and Technology (NIST),
k{Juliet Test Suite v1.2 for C/C++}, 
 PDF: \path{https://samate.nist.gov/SRD/resources/Juliet_Test_Suite_v1.2_for_C_Cpp_-_User_Guide.pdf},
 Zip File: \path{https://samate.nist.gov/SRD/testsuites/juliet/Juliet_Test_Suite_v1.2_for_C_Cpp.zip}.

\bibitem{parr:patterns}
T. Parr,
\newblock{Language Implementation Patterns},
\newblock {Pragmatic Bookshelf}, 2010.

\bibitem{peachfuzzer}
Peach fuzzing platform,
\newblock {} 
\newblock {\path{http://peachfuzzer.com/}}

\bibitem{intflow:hybrid}
M. Pomonis, T. Petsios, K. Jee, M. Polychronakis, and A. D Keromytis,
\newblock{IntFlow: Improving the Accuracy of Arithmetic Error Detection Using Information Flow Tracking},
In \textit {Proceedings of the Annual Computer Security Applications Conference (ACSAC)}, 2014.

\bibitem{benefits}
D. M. Rafi, K. Moses, K. Petersen, and Mika V. M{\"a}ntyl{\"a},
\newblock{Benefits and limitations of automated software testing: systematic literature review and practitioner survey},  
In \textit {Proceedings of the 7th International Workshop on Automation of Software Test (AST)}, 2012.

\bibitem{early:repairs}
B.-C. Rothenberg and O. Grumberg,
\newblock{Sound and Complete Mutation-Based Program Repair},  
In \textit { the International Symposium on Formal Methods (FM)}, 2016.

\bibitem{integerlib}
R.C. Seacord,
\newblock{The CERT C Secure Coding Standard, Addison-Wesley Professional}, 2008.

\bibitem{safeint}
SafeInt, 
 \path{http://safeint.codeplex.com/}.

\bibitem{fix1}
CVE-2017-7975, 
 \path{https://nvd.nist.gov/vuln/detail/CVE-2017-7975#vulnDescriptionTitle}.
 See repair: \path{https://goo.gl/Spx5Qv}.

\bibitem{fix2}
CVE-2016-10164, 
 \path{https://access.redhat.com/security/cve/cve-2016-10164}.
 See repair: \path{https://goo.gl/ufnNrZ}.

\bibitem{fix3}
CVE-2016-10164, 
 \path{https://access.redhat.com/security/cve/cve-2016-8706}.
 See repair: \path{https://goo.gl/Hrh39i}.

\bibitem{fix4}
CVE-2016-9427, 
 \path{https://cve.mitre.org/cgi-bin/cvename.cgi?name=CVE-2016-9427}.
 See repair: \path{https://goo.gl/Tqqmi3}.

\bibitem{fix5}
CVE-2014-9862, 
 \path{https://nvd.nist.gov/vuln/detail/CVE-2014-9862}.
 See repair: \path{https://goo.gl/iKodLf}.

\bibitem{monirul}
M. I Sharif, A. Lanzi, J. T. Giffin, and W. Lee,
\newblock {Impeding malware analysis using conditional code obfuscation},
In \textit{Proceedings of the Network and Distributed System Security Symposium (NDSS)}, 2008.

\bibitem{targeted}
S. Sidiroglou-Douskos, E. Lahtinen, N. Rittenhouse, P. Piselli, F. Long, D. Kim, and M. Rinard,
\newblock {Targeted Automatic Integer Overflow Discovery Using Goal-Directed Conditional Branch Enforcement},
In \textit {Proceedings of the International Conference on Architectural Support for Programming Languages and Operating Systems (ASPLOS)}, 2015.

\bibitem{tap}
S. Sidiroglou-Douskos, E. Lahtinen, and M. Rinard,
\newblock {Automatic Discovery and Patching of Buffer and Integer Overflow Errors},
 \path{https://dspace.mit.edu/handle/1721.1/97087},
In \textit {MIT-CSAIL-TR-2015-018}, 2015.

\bibitem{codephage}
S. Sidiroglou-Douskos, E. Lahtinen, F. Long, and M. Rinard,
\newblock{Automatic error elimination by multi-application code transfer}, 
In \textit {Proceedings of the International Conference on Architectural Support for Programming Languages and Operating Systems (ASPLOS)}, 2015.

\bibitem{inteq}
H. Sun, X. Zhang, Y. Zheng, and Q. Zeng,
\newblock {IntEQ: Recognizing Benign Integer Overflows via Equivalence Checking Across Multiple Precisions},
In \textit {Proceedings of the International Conference on Software Engineering (ICSE)}, 2016.

\bibitem{nvd}
United States National Vulnerability Database (NVD),
\newblock {} 
 \path{https://nvd.nist.gov/vuln/search}.

\bibitem{wang:intscope}
T. Wang, T. Wei, Z. Lin, and W. Zou,
\newblock {IntScope: Automatically Detecting Integer Overflow Vulnerability in x86 Binary Using Symbolic Execution},
In \textit {Proceedings of the Network and Distributed System Security Symposium (NDSS)}, 2009.

\bibitem{taintscope}
T. Wang, T. Wei, G. Gu, and W. Zou,
\newblock{TaintScope: A checksum-aware directed fuzzing tool for automatic software vulnerability detection},  
In \textit {Proceedings of the Symposium on Security and Privacy (S\&P)}, 2010.

\bibitem{kint}
X. Wang, H. Chen, Z. Jia, N. Zeldovich, and M.F. Kaashoek,
\newblock{Improving integer security for systems with KINT}, 
In \textit {Proceedings of the USENIX Security Symposium (USENIX SEC)}, 2012.

\bibitem{soupint}
T. Wang, C. Song, and W. Lee,
\newblock {Diagnosis and Emergency Patch Generation for Integer Overflow Exploits},
In \textit {Proceedings of the International Conference on Detection of Intrusions and Malware, and Vulnerability Assessment (DIMVA)}, 2014.

\bibitem{uqbtng:wojtczuk}
R. Wojtczuk,
\newblock{UQBTng: A Tool Capable of Automatically Finding Integer Overflows in Win32 Binaries},

\bibitem{herley:security:science}
C. Herley, and P. C. van Oorschot,
\newblock {Science, Security, and the Elusive Goal of Security as a Scientific Pursuit},
In \textit {Proceedings of the International Symposium on Research in Attacks, Intrusions, and Defenses (RAID)}, 2015.

\bibitem{coverity}
Coverity Scan Static Analysis,
 \path{https://scan.coverity.com/}.

\bibitem{c++:standard}
Working Draft, Standard for Programming Language C++
\newblock{N4296},  
 \path{http://www.open-std.org/jtc1/sc22/wg21/docs/papers/2014/n4296.pdf}.

\bibitem{integer:binary}
Y. Zhang, X. Sun, Yi Deng, L. Cheng, S. Zeng, Y. Fu, and D. Feng,
\newblock {Improving Accuracy of Static Integer Overflow Detection in Binary},
In \textit {Proceedings of the International Symposium on Research in Attacks, Intrusions, and Defenses (RAID)}, 2015.

\bibitem{int:patch}
Q. Xiao, Y. Chen, H. Huang, L. Qi,
\newblock {IntPatch: Automatically Fix Integer-Overflow-to-Buffer-Overflow Vulnerability at Compile-Time},
In \textit {Proceedings of the European Symposium on Research in Computer Security (ESORICS)}, 2010.

\bibitem{toolx}
B. Zhang, C. Feng, B. Wu, and C.Tang,
\newblock {Detecting Integer Overflow in Windows Binary Executables based on Symbolic Execution},
In \textit {Proceedings of the International Conference on Software Engineering, Artificial Intelligence, Networking and Parallel/Distributed Computing (SNPD)}, 2016.


\bibitem{khedker:flow}
U. P. Khedker, A. Sanyal, and B. Karkare,
\newblock{Data Flow Analysis},
\newblock {CRC Press}, 2009.

\bibitem{sharir:pnueli}
M. Sharir and A. Pnueli,
\newblock{Two Approaches to Interprocedural Data Flow Analysis},
\newblock {Program Flow Analysis: Theory and Applications}, Prentice-Hall, 1981.

\bibitem{tip:slicing}
F. Tip,
\newblock {A Survey of Program Slicing Techniques},
\newblock {In Journal of Programming Languages},
 \path{http://www.franktip.org/pubs/jpl1995.pdf}, 1995.

%
%
%
%
%
%
%
%
%
%

\bibitem{coker:integers}
Z. Coker, and M. Hafiz,
\newblock {Program Transformations to Fix C Integers},
In \textit {Proceedings of the International Conference on Software Engineering (ICSE)}, 2013.
%
%

%
%
%
%
%
%
%
%

%
%



\end{thebibliography}


\end{document}